\documentclass[12pt,preprint]{aastex}

\newcommand{\beq}{\begin{equation}}
\newcommand{\eeq}{\end{equation}}

\def\ln{{\rm ln}}

\def\3he{$^3$He\,}
\def\he4{$^4$He\,}

\shorttitle{Electron Acceleration}
\shortauthors{Liu et al.}

\begin{document}

\title{Elementary Energy Release Events in Flaring Loops: Effects of
Chromospheric Evaporation on X-rays} 

\author{
Siming Liu\altaffilmark{1}, Feiran
Han\altaffilmark{2}, and Lyndsay Fletcher\altaffilmark{1}
}
\altaffiltext{1}{Department of Physics and Astronomy, University of
Glasgow, Glasgow, G12 8QQ, UK; sliu@astro.gla.ac.uk}
\altaffiltext{2}{Purple Mountain Observatory, Chinese Academy of
Sciences, Nanjing 210008, China}

\begin{abstract}

 With the elementary energy release events introduced in a previous
 paper \citep{l09} we model the chromospheric evaporation in flaring
 loops. The thick-target hard X-ray (HXR) emission produced by
 electrons escaping from the acceleration region dominates
 the impulsive phase and the thin-target emission from the acceleration
 region  dominates 
 the low-energy thermal component in the gradual phase, 
 as observed in early impulsive flares. Quantitative details depend on
 properties of the thermal background, which leads to
 variations in the correlation between HXR flux and spectral index.
 For lower temperature and/or higher density of the background
 electrons, the HXRs both rise and decay more quickly with a plateau
 near the peak. The plateau is less prominent at higher energies.
 Given the complexity of transport
 of mass, momentum, and energy along loops in the impulsive phase, we
 propose a  strategy to apply this  
 single-zone energy release and electron acceleration model to
 observations of flares associated with single loops so that the
 energy release, electron acceleration, and evaporation processes may be 
 studied quantitatively.

\end{abstract}


\section{INTRODUCTION}
\label{intro}


The presence and nature of accelerated particles in solar flares are
deduced primarily from secondary impulsive radiations. Because of the
short thermalization and radiative timescales in the lower solar
atmosphere, low-energy impulsive
emissions from the dense chromosphere to photosphere can be thermal as
well. Gradual emission components are presumably thermal and early in
the flare are strongly
linked to the chromospheric evaporation. The electron acceleration
timescale is much shorter than that for the
chromospheric evaporation
\citep{m97, a02}, and it appears that the two processes are decoupled.
However, electron acceleration appears to be an 
inevitable component of the flare energy release \citep{k08}, which
itself is multi-scale as 
evident from the rich temporal, spatial, and energetic characteristics of
flares, and electron
acceleration in flares 
can be very efficient. Solar
flares are caused by energy release on macroscopic scales, and the
amount of nonthermal electrons inferred from observations can be a significant
fraction of the background electrons \citep{bm77, f08, k09s}.  Energetic
electrons in flares therefore are mostly accelerated 
from the background plasma, and the acceleration should depend in part
on the background properties \citep{p04, l09, l09w}. One therefore may
probe this dependence with observations over a relatively long
timescale, when the properties of the background plasma have changed
significantly. This is especially true for flares dominated by single
loops \citep{v05, l06, x08, h08}, where the modification of the
background plasma through chromospheric evaporation is expected to affect the
characteristics of nonthermal radiations.

In the context of stochastic particle acceleration by turbulent
electromagnetic fields, low-energy particles mostly couple with each other
through Coulomb collisions. In the high-energy range the energy gain
of particles through interactions with the turbulent fields dominates
over the energy loss through Coulomb collisions with lower energy particles
and radiation \citep{b77}. A power-law high-energy particle
distribution forms as this energy gain competes with the particle escape from 
the acceleration region through spatial diffusion with the power-law
spectral index determined by the acceleration and escape rates.
The particle 
distribution in the acceleration region  generally has a thermal
form at low energies and a high-energy power-law tail with the
transition from the thermal to nonthermal component located at the
energy $E_t$, where the particle energy gain rate is comparable to
the energy loss rate. Over the past few decades, it has been
demonstrated that this mechanism is compatible with many flare observations
\citep{h92, m95, p04}. However, there are very few testable
predictions from detailed modeling. The same is true for other particle
acceleration mechanisms \citep{h85, m97, t98, a02, d06}.

Given the important roles that turbulent electromagnetic fields play
in the energy release processes of solar flares, in a previous paper we
pointed out that direct modeling of the dynamics of turbulence may
advance our studies of flares and lead to testable predictions
\citep{l09}. In the simplest scenario, where the turbulence is generated in a
closed loop with a characteristic intensity and length scale --- an
elementary energy release event --- the
emission characteristics of the flaring loop are relatively well
determined. We showed that these elementary energy release events
might explain the soft-hard-soft spectral evolution of hard X-ray (HXR)
pulses observed frequently both in chromospheric and coronal sources
\citep{b06}. In this 
paper, we have a summary of the elementary energy release events
(Section \ref{events}) and 
quantify the chromospheric evaporation and X-ray emissions driven by this energy
release (Section \ref{model}). 
The model 
predicts faster rise and decay of HXR fluxes with a plateau near the
turbulence peak for higher values of the ratio of $E_t$ to the thermal
energy at this peak. The correlation between the nonthermal HXR flux
and spectral index also depends on the temperature and/or density of the
background electrons (Section \ref{results}). This simple model
does not quantitatively address the plasma heating and many important
transport processes in 
flaring loops. We show, however, that it is possible to use RHESSI
observations to test the essence of the model and quantify the energy
release, particle acceleration, and evaporation processes (Section
\ref{dis}). Conclusions are drawn in Section \ref{con}.

\section{Elementary Energy Release Events}
\label{events}

Elementary flare
bursts were noticed in early spectral observations of HXR flares, where
these bursts are well correlated among all available energy channels \citep{v74,
d78}. Later high-resolution observations reveal soft-hard-soft spectral
evolution, which \citet{g04} attribute to elementary acceleration events.
Although the duration and peak flux may vary significantly from burst
to burst, these events can be readily identified by the soft-hard-soft
behavior.\footnote{Decomposition of HXR light curves into elementary
bursts is not always possible, especially for `extended bursts', whose
spectral evolution can be quite complicated
\citep{h76, kh08, g08, s09}.}
Theoretically we called them elementary energy release events. 
The model for the elementary energy release event and stochastic
acceleration of electrons is described in \citet{l09}. For the sake of
completeness, we summarize the key results here.
The energy release is characterized with a turbulence generation length
scale $l_e$ and  turbulence energy density $b_0^2 B^2/4\pi$, where $B$
is the background magnetic 
field and $b_0$ is a dimensionless parameter characterizing the fraction
of magnetic energy dissipated during the event. Considering the
energy equipartition between the turbulence kinetic and magnetic
energies, the eddy speed at $l_e$ is given by 
$v_{e0}=b_0 v_{\rm A}$, where $v_{\rm A}= B/(4\pi\rho)^{1/2}$ is the 
Alfv\'{e}n speed and $\rho$ is the background mass density. The
characteristic timescale is then given by $\tau_r = l_e/v_{e0}$. 

Detailed analysis of correlation between HXR flux and spectral index
reveals complicated quantitative results \citep{g04, g06, g08}. The shapes of
these bursts also show significant variations \citep{s07, ly09}. These
variations can be caused by changes in properties of the background
plasma as the energy release triggers responses in coronal and
chromospheric plasmas. The topology of magnetic fields may
introduce additional features. We focus on flares associated with single
loops so that particle acceleration, plasma heating, and
chromospheric evaporation are the dominant physical processes in the
impulsive phase.

In the 1D model for turbulence cascade and stochastic particle
acceleration by \citet{m95}, the particle acceleration and plasma
heating start almost instantaneously upon the arrival of turbulence at
certain small
scales, where background particles resonate effectively with turbulent plasma
waves. Applying such a model to elementary energy events would imply almost
universal rise profile of HXR bursts. There is little observational
evidence for such a universal HXR rise profile, and the turbulence
cascade and particle 
acceleration are truly 2D processes, which are highly nonlinear,
complicated, and not well understood \citep{j09}. Here we use $b B$ with
$b\le b_0$ to characterize the effective intensity of turbulent magnetic
fields in resonance with background particles and 
assume that
\begin{equation}
\dot{b}(t)/b(t) =\left\{
\begin{array}{ll}
\tau_r^{-1} = b_0v_{\rm A}/l_e\ \ \ \ &{\rm for\ the\ rise\ phase}\,, \\
-\tau_d^{^-1}= -b^2v_{\rm A}/l_e\ \ \ \ &{\rm for\ the\ decay\ phase}\,, \\
\end{array}
\right.
\label{dotd}
\end{equation}
where the `dot' 
indicates a derivative with respect to time
$t$ and the Kraichnan phenomenology has been used to obtain the
turbulence decay time $\tau_d$ \citep{k65, m95}.
Both the kinetic and magnetic energy density of this turbulence is given by
$b^2 \rho v_{A}^2/2$. The energy dissipation rate is then given by
\begin{equation}
\dot{\Sigma}(t) \equiv 2 \rho v_{\rm A}^2 b^2/\tau_d= 2\rho v_{\rm
 A}^2b^4v_{\rm A}/l_e\,. 
\label{energy}
\end{equation}
Note that $\rho v_{\rm A}^2= B^2/4\pi$ does not change with time and the
derivative of $b^2$ is $2b\dot{b}$.
There are also uncertainties of factors of a few in these timescales.

As in the standard quasi-linear theory, the wave-particle interaction
rates are proportional to the effective turbulence energy density
$b^2\rho v_{\rm A}^2$. In cases where the particle distribution is
isotropic, the pitch-angle averaged electron 
acceleration and scattering timescales by the turbulence can then be
parameterized, respectively, with 
\begin{eqnarray}
\tau_{ac}=S_a \tau_t b^{-2}\,, \ \ \ \ \ 
\tau_{sc}=S_s \tau_t b^{-2}\,,
\label{tac}
\end{eqnarray}
where $\tau_t=l_e/v$, $v$ is the electron speed, and $S_a$ and $S_s$ are
two dimensionless coefficients characterizing 
the acceleration and scattering of electrons by the turbulent fields. We
use a leaky-box loss term to account for the spatial diffusion of
electrons along magnetic field lines. The corresponding
electron escape time from the acceleration region  
\begin{equation}
\tau_{esc} = l_0^2/(v^2\tau_{sc})+l_0/(\sqrt{2}v)=
 [l_0b^2/(l_eS_s)+2^{-1/2}]l_0/v\,, 
\label{esc}
\end{equation}
where $l_0>l_e$ is the length of the energy release site. 
This equation is slightly different from that
in \citet{l09}. Considering that the escape time cannot be shorter
than the mean transit time $l_0/(\sqrt{2}v)$, we include the second term on the
right-hand side. This term is important only in the limit of weak
scattering when the scattering time $\tau_{sc}$ is much longer than
$\sim l_0/v$ \citep{p99}.  
In a more accurate treatment, one also needs to consider the reduction
of the scattering time by Coulomb collisions \citep{p04}. The trapping
of electrons by magnetic loops converging at both footpoints will also
increase the escape time \citep{m95}. In the regime, where
scattering by turbulence dominates, the effects of Coulomb scattering, the
transit-time term, and magnetic trapping may be negligible and one
obtains the expression in \citet{l09}.

In general, the electron distribution function at a given time $t$
depends on both the 
electron momentum and spatial location. For isotropic distributions, the
spatially integrated electron distribution in the acceleration region
only depends on the electron kinetic energy $E$:  
$N(E, t)=N_0(t) g(E, t)$, where $N_0=Al_0 n$ is the total number of free
electrons in the acceleration region , $n\propto \rho$ is the average
electron density, and $A$ is the cross section of 
the flaring loop. We first ignore the structure of the flaring loop and
assume that the evolution timescale of the distribution function is much
shorter than $\tau_r$. Then 
the normalized quasi-steady distribution function $g(E, t)$ can be heuristically
prescribed as:
\begin{equation}
g(E, t) = {g_0(E_t/k_{\rm B}T_e, \delta)\over (k_{\rm B}T_e)^{3/2}}\left\{
\begin{array}{ll} 
E^{1/2}\exp({-E/k_{\rm B} T_e}) & \textrm{for $E<E_t$}\,,\\
E_t^{1/2}\exp({-E_t/k_{\rm B}T_e})(E/E_t)^{-\delta} & \textrm{for
 $E\ge E_t$}\,, 
\end{array}  \right.
\end{equation}
where 
\begin{equation}
\delta = 1/2+(1+\tau_{ac}/4\tau_{esc})^{1/2} =
 1/2+\{1+S_aS_sl_e^2/[2^{3/2}b^2 l_0(l_eS_s+2^{1/2}b^2l_0)]\}^{1/2}\,,
\label{gam}
\end{equation}
\begin{equation}
g_0(r, \delta) =
 [r^{3/2}\exp(-r)/(\delta-1)-r^{1/2}\exp(-r)+\pi^{1/2}
 \mbox{erf}(r^{1/2})/2]^{-1}\,,   
\end{equation}
$g_0$ is obtained from $\int g(E){\rm d} E=1$ and the time dependence is
carried through $T_e(t)$, $E_t(t)$, and $\delta(t)$. \footnote{A more accurate
treatment of $g(E)$ with the effects of Coulomb collisions taken
into account can be found in \citet{g05}, and considering the loop structure
will lead to a multi-thermal distribution, which may mimic a power law.}
The transition energy between the thermal and power-law components 
\begin{equation}
E_t =
(2\pi\ln\Lambda S_a l_e n)^{1/2} 
e^2/b \,
\label{et}
\end{equation}
is obtained from the equality of $E/\tau_{ac}(E)$ to the
Coulomb energy loss rate 
\begin{equation}
\dot{E}_c(E, t)=2^{3/2}\pi\ln\Lambda e^4 n(t)/(Em)^{1/2}
\label{coul}
\end{equation} 
at
$E_t$, where $T_e$, 
$m$, $e$, $k_{\rm B}$, and  $\ln\Lambda\simeq 20$ are 
the electron 
temperature, mass, charge, Boltzmann constant, and
Coulomb logarithm, respectively:
\begin{equation}
2^{1/2} E_t^{3/2} b^2/(S_a l_e m^{1/2}) =2^{3/2}\pi\ln\Lambda e^4
n/(E_tm)^{1/2}.
\label{tran} 
\end{equation}
 The dependence of $E_t$ on $b$ and $n$
comes from the acceleration time and Coulomb energy loss rate, respectively.
Then the (thermally) normalized transition energy 
\begin{equation}
r(t)\equiv E_t(t)/k_{\rm B}T_e(t)=
r_0[n(t)/n_0]^{1/2} [b_0/b(t)] 
[T_{e0}/T_e(t)]\,
\end{equation} 
where, and in what follows, the subscript `0' indicates
quantities at the peak of $b(t)$. The dependence of $r(t)$ on $T_e$ is
due
to energy normalization to the thermal energy $k_{\rm B}T_e$. 

For convergence of the energy flux carried away by escaping
electrons, the index of $g(E)/\tau_{esc}$, $\delta-1/2$, must be 
greater than 2 (i.e., $\delta> 5/2$) if there is no high-energy
cutoff in $g(E)$. From equations (\ref{gam}), 
(\ref{tac}), and (\ref{esc}),
we then have 
\begin{equation}
\tau_{ac}>12\tau_{esc}\,, \, \ \ {\rm}\  \ \
 b^2<[(1+2S_a/3S_s)^{1/2}-1]l_eS_s/(2^{3/2}l_0) 
\equiv \Delta^2\,
\,.
\label{Del}
\end{equation}
$\Delta$ is an upper limit for $b_0$. As $b_0$ approaches to $\Delta$,
the energy flux carried away by nonthermal electrons increases
quickly near the turbulence peak. The damping of turbulence by nonthermal
electrons will be 
important and equation (\ref{dotd}) needs to be modified accordingly. 
These complex nonlinear effects can be considered wherever there are
sufficient theoretical and/or observational justifications
\citep{bf09}. There is also a physical upper limit on $\Delta$. 
Since the turbulence energy density $b_0^2 \rho v_{\rm A}^2$ should be
less than the magnetic field energy density $B^2/8\pi$, we have $b_0<
\Delta < 2^{-1/2}$. For $b_0\ge 2^{-1/2}$, the flaring loop will not be
well defined and the turbulence pressure may drive the
acceleration site into an expansion. The model needs to be modified
properly to account 
for these kinds of events. Therefore $b_0<
\Delta < 2^{-1/2}$ defines a self-consistent domain of the model. 

The escape of thermal electrons
from the acceleration region  can be treated as conduction. However,
scattering by turbulence will suppress the spatial diffusion of electrons near
$E_t$, which will lead to a conductivity lower than the Spitzer
conductivity \citep{j06}. The loop structure and flow motions may also
affect the conduction significantly \citep{a78}. In this paper, we do
not seek more accurate 
treatment of electron transport at low energies. The formulae above is
assumed to be applicable in the whole energy range. 

From equations (\ref{gam}) and (\ref{Del}), we then have 
\begin{equation}
b^2 =
{\left\{(1+2S_a/[(\delta-0.5)^2-1)S_s]^{1/2}-1\right\}l_eS_s\over 2^{3/2}l_0} =
\Delta^2\xi^2(\delta, S_a/S_s)
\end{equation}
where
\begin{equation}
\xi^2(\delta, s)=
{\left\{1+2s/[(\delta-0.5)^2-1]\right\}^{1/2}-1\over
(1+2s/3)^{1/2}-1}<1\,. 
\end{equation}
One can show that $\xi(\delta, s)<1$ is equivalent to $\delta>5/2$.
It follows
\begin{eqnarray}
r
&=&
{r_0b_0n(t)^{1/2}T_{e0}\over \Delta \xi(\delta, S_a/S_s) n_0^{1/2}T_e(t)}
\,,\\
\delta &=&
 {1\over 2}+\left\{1+{S_a/S_s\over
 b^4\Delta^{-4}[1+S_a/3S_s-(1+2S_a/3S_s)^{1/2}] + 
 b^2\Delta^{-2}[(1+2S_a/3S_s)^{1/2}-1]}\right\}^{1/2} \,.
\label{gam1}
\end{eqnarray}
For $S_a\gg S_s$, $\xi^2(\delta, s)
= \left\{3/[(\delta-0.5)^2-1]\right\}^{1/2}$, $\delta =
1/2+(1+3\Delta^4/b^4)^{1/2}$, and $\Delta^2=(S_aS_s/12)^{1/2}(l_e/l_0)$
in agreement with \citet{l09}. 

\section{X-rays from a Flaring Loop and Chromospheric Evaporation}
\label{model}

If we assume thin-target emission from the acceleration region (coronal
sources) and  
thick target for escaping electrons (chromospheric footpoints),
the overall {\it specific intensity} of photons observed at Earth is
\begin{equation}
I(\varepsilon, t)={\varepsilon\over 2^{3/2}m^{1/2}\pi R^2}
 \int_\varepsilon^\infty\left\{N(E, t)+{1\over \dot{E}_c(E, t)}\int_{E}^\infty 
{N(E^\prime, t)\over \tau_{esc}(E^\prime, t)}{\rm d} E^\prime\right\}
 E^{1/2} n(t) \sigma(E, \varepsilon){\rm d} E\,,
\label{I}
\end{equation} 
where $R=1$ AU, $\varepsilon$, 
and $\sigma(E,\varepsilon)$
are, respectively, the distance between 
the Sun and Earth, the photon energy,  
and the
angle-integrated bremsstrahlung emission cross section \citep{b34}. The first
and second terms on the right hand side correspond to
the thin-target acceleration region  and thick-target footpoints, respectively,
 and transport effects from the acceleration region  to the
 thick-target region have 
 been ignored. $E_c(E, t)$ and $\tau_{esc}(E, t)$ are given by equations
 (\ref{coul}) and (\ref{esc}), respectively \citep{p99}. One can show
 that 
$
\dot{E}_c(E, t)\tau_{esc}(E^\prime, t) = \dot{E}_c(E_t,
 t)\tau_{esc}(E_t, t){E_t(t)/(EE^\prime)^{1/2}} =
 {E_t^2(t)\tau_{esc}(t)/[ (EE^\prime)^{1/2}\tau_{ac}(t)]}\,,  
$
where we have used equation (\ref{tran}). From equation (\ref{gam}), we
have $\tau_{esc}(t)/\tau_{ac}(t) = 4/\{[\delta(t)-1/2]^2-1\}$. It follows
that 
\begin{equation}
\dot{E}_c(E, t)\tau_{esc}(E^\prime, t) = {4[r(t)k_{\rm B}T_e(t)]^2\over
 \{[\delta(t)-1/2]^2-1\} (EE^\prime)^{1/2}}\,.
\end{equation}
Therefore $I(\varepsilon, t)$ only depends on $N(E, t)$ and $n(t)$,
and the relation between the thin- and thick-target component is
determined by $N(E, t)$ alone, which itself is determined by $r(t)$,
$n(t)$, and $T(t)$. 

For elementary energy release events, equation (\ref{dotd}) governs the
evolution  of 
$b(t)$, which can be used to derive $E_t(t)$ and $\delta(t)$. One also needs
to know the evolution of $N_0(t)$, $T_e(t)$, and 
$n(t)$ to obtain the evolution of $N(E, t)$, which
dictates the radiative 
characteristics. For events associated with
closed loops, one may assume that $l_0$ and $A$ 
do not
change with time. $N_0(t)$ is then proportional to the average density
$n(t)$. In principle, one can 
solve the mass and energy conservation equations for the plasma in the flaring
loops to study the detailed hydrodynamical response of the loops to the
energy deposition \citep{k06, l09w}. Here we are mostly interested in the
spatially integrated properties. So a much simpler
treatment is possible. 

One inevitable consequence of the impulsive energy
deposition into closed coronal loops is chromospheric evaporation.
This evaporation can be caused through thermal conduction
\citep[e.g.,][]{a78},  nonthermal 
particles \citep[e.g.,][]{b73}, and wave or turbulence dissipation at the
footpoints 
\citep{f08, bt09, h09}. The 
details of these processes are likely to be complicated and may affect
low-energy emission characteristics significantly \citep{k06}. For
studies of X-rays, these processes may introduce short timescale
features, which are energetically less important. This paper focuses on the
dominant X-ray emission component on a timescale of $\tau_r=l_e/b_0 v_{\rm
A}$. For the sake of simplicity, we assume that the average density
increase rate in the coronal loop is proportional to the energy
dissipation rate  
$\dot{\Sigma}$, i.e., 
\begin{equation}
\dot{n}(t) = C_en_i\dot{\Sigma}(t)/(\rho v_{\rm A}^2) = 2C_en_i b^4v_{\rm
 A}/l_e\,, 
\label{evap}
\end{equation}
where $C_e$ is a dimensionless coefficient and the subscript `$i$'
indicates quantities at the flare onset as is in what follows. Note that
$\rho v_{\rm A}^2=B^2/4\pi$ gives the normalization of the energy density and
does not change with time. $n_i$ can be considered as a model parameter
to be determined by observations.

The above energy conservation equation can be generalized as follows
\begin{equation}
3 k_{\rm B}[n(t)\dot{T}(t)+\dot{n}(t)T(t)] = 3 k_{\rm B}C_e
 T(t)n_i[\dot{\Sigma}(t)-\Lambda(t)]/(\rho v_{\rm A}^2)\,,
\label{evap3}
\end{equation}
where we have assumed that the electron and ions temperature are equal
and the internal energy density is given by $3nk_{\rm B}T$ with $T(t)=T_e(t)$.
$\Lambda(t)$ represents the overall energy loss rate from the flaring loop.
If energies in forms other than the background magnetic field,
turbulence, and internal energies of the thermal plasma 
are negligible, we have  $3 k_{\rm B}[n(t)\dot{T}(t)+\dot{n}(t)T(t)]
=\dot{\Sigma}(t)-\Lambda(t)$, which gives
\begin{equation}
C_e = \rho v_{\rm A}^2/(3 k_{\rm B} T n_i)\,.  
\label{evap4}
\end{equation}
Therefore $C_eT$ should be approximately a constant and we recover
equation (\ref{evap}) for $\dot{T}=0$ and $\Lambda = 0$.

Equations (\ref{dotd}) and (\ref{evap}) then give  
\begin{equation}
n(t) =n_i\left\{
\begin{array}{ll}
1+C_e [b(t)^4-b_i^4]/(2b_0) \equiv a^2 + C_e b^4/(2 b_0)\ \ \ \ &{\rm for\
 the\ rise\ phase}\,, \\ 
a^2+C_e b_0^3/2+C_e[b_0^2-b^2(t)]\ \ \ \ 
&{\rm for\ the\ decay\ phase}\,. \\
\end{array}
\right.
\label{density}
\end{equation}
Combining equations
(\ref{density}) and (\ref{dotd}), we have for the rise phase
\begin{equation}
4b_0 t v_{\rm A0}/l_e =\left\{
\begin{array}{ll}
h(t)
+ a[n_i/n_0]^{1/2}\
 \ln\left\{{[n(t)^{1/2}-a{n_i}^{1/2}][n_0^{1/2}+a{n_i}^{1/2}]\over
 [n(t)^{1/2}+a{n_i}^{1/2}][n_0^{1/2}-a{n_i}^{1/2}]}\right\} 
\ &{\rm for\ } 1>C_e b_i^4/(2b_0)\,, \\
2(b^2-b_0^2)/b_0^2= h(t) \ \ &{\rm for\ }1=C_e b_i^4/(2b_0)\,, \\
h(t)-2a_1[n_i/n_0]^{1/2}\{{\rm
 tg}^{-1}[n/(n_ia_1^2)]^{1/2}-{\rm tg}^{-1}[n_0/(n_i a_1^2)]^{1/2}\}
\ \ &{\rm for\ }1<C_e b_i^4/(2b_0)\,, \\
\end{array}
\right.
\label{b}
\end{equation}
where $a_1 =[C_e b_i^4/(2b_0)-1]^{1/2}$, 
$v_{{\rm A}0} = B/(4\pi\rho_0)^{1/2}$, 
and $h(t)=2[n(t)^{1/2}-n_0^{1/2}]/n_0^{1/2}$. 
From the second expression, we
see that the flare starts after $t_s= -l_e/(2v_{{\rm A}0}b_0)$ when
$b(t_s)=0$ and $n(t_s)=0$. The third expression can also be extrapolated
backward in time to 
a point $t_s$ when $n(t_s)=0$, i.e., $C_e b_i^4/(2b_0)=1+C_eb_s^4/(2b_0)$,
where $b_s = b(t_s)$.  
Note that $n_0/n_i =
a^2+C_eb_0^3/2=C_eb_0^3/2-a_1^2 = 1+C_e(b_0^4-b_i^4)/(2b_0)$, and from the
third expression we have 
\begin{eqnarray}
2b_0 t_s v_{\rm A0}/l_e & = &-1+\left[{C_e
b_i^4/(2b_0)-1\over 1 + 
 C_e(b_0^4-b_i^4)/(2b_0)}\right]^{1/2}{\rm tg}^{-1}{\left[1 +
 C_e(b_0^4-b_i^4)/(2b_0)\over C_e
b_i^4/(2b_0)-1\right]^{1/2} }\, \nonumber \\
 & = &-1+
[(b_0/b_s)^4-1]^{-1/2}{\rm tg}^{-1}{\left[
 (b_0/b_s)^4-1\right]^{1/2} }\, .
\end{eqnarray}
The solution for the decay
phase can be obtained numerically.

We consider 
first the solution in the middle
of equation (\ref{b}). Then 
$C_e = 2b_0/b_i^4 = 2n_0/(n_i b_0^3)$, and
$n=n_0b^4/b_0^4$ and $n_0(1+2/b_0-2b^2/b_0^3)$ for the rise and decay
phase, respectively. 
If we assume a coronal background level for
$b_i$ and $n_i$, the density increases by a factor of $b_0^4/b_i^4$
and $1+2/b_0$ in the rise and decay phase, respectively. There is more
evaporation for stronger events especially during the rise phase. From
equation (\ref{evap}), one has the
energy dissipation in the rise and decay phase of
$(1-b_i^4/b_0^4)(b_0^3/2)\rho v_{\rm A}^2$ and $b_0^2\rho v_{\rm A}^2$,
respectively. The energy dissipated in the decay phase is equal to the
total turbulence energy at the peak as expected. In the rise phase,
equation (\ref{dotd}) shows that there is a continuous energy deposition
into the loop and the energy dissipation depends on the evaporation
process. 
  
\section{Results}
\label{results}

\begin{figure}[ht]
\begin{center}
 \includegraphics[width=5.5cm]{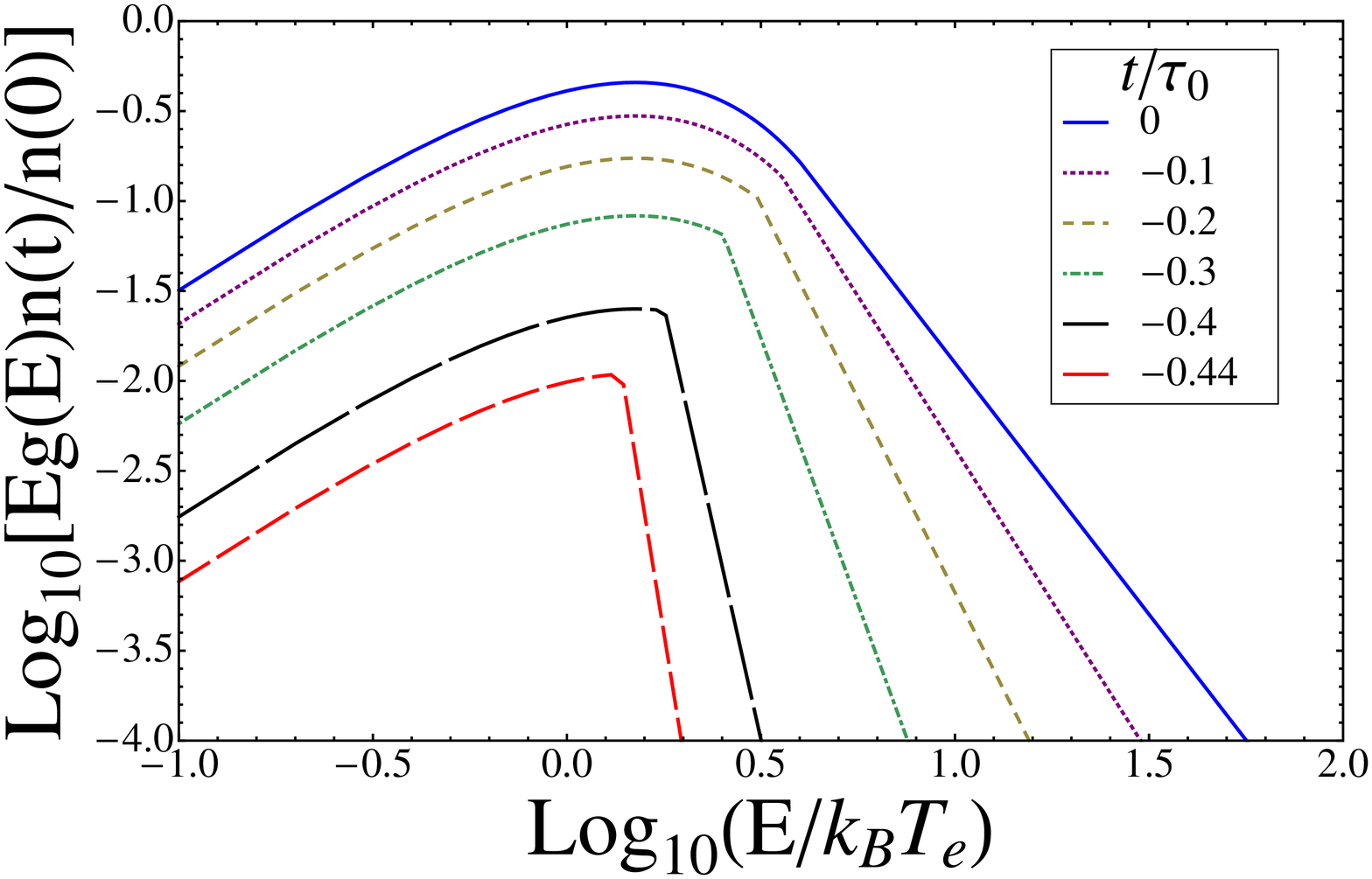}
\includegraphics[width=5.5cm]{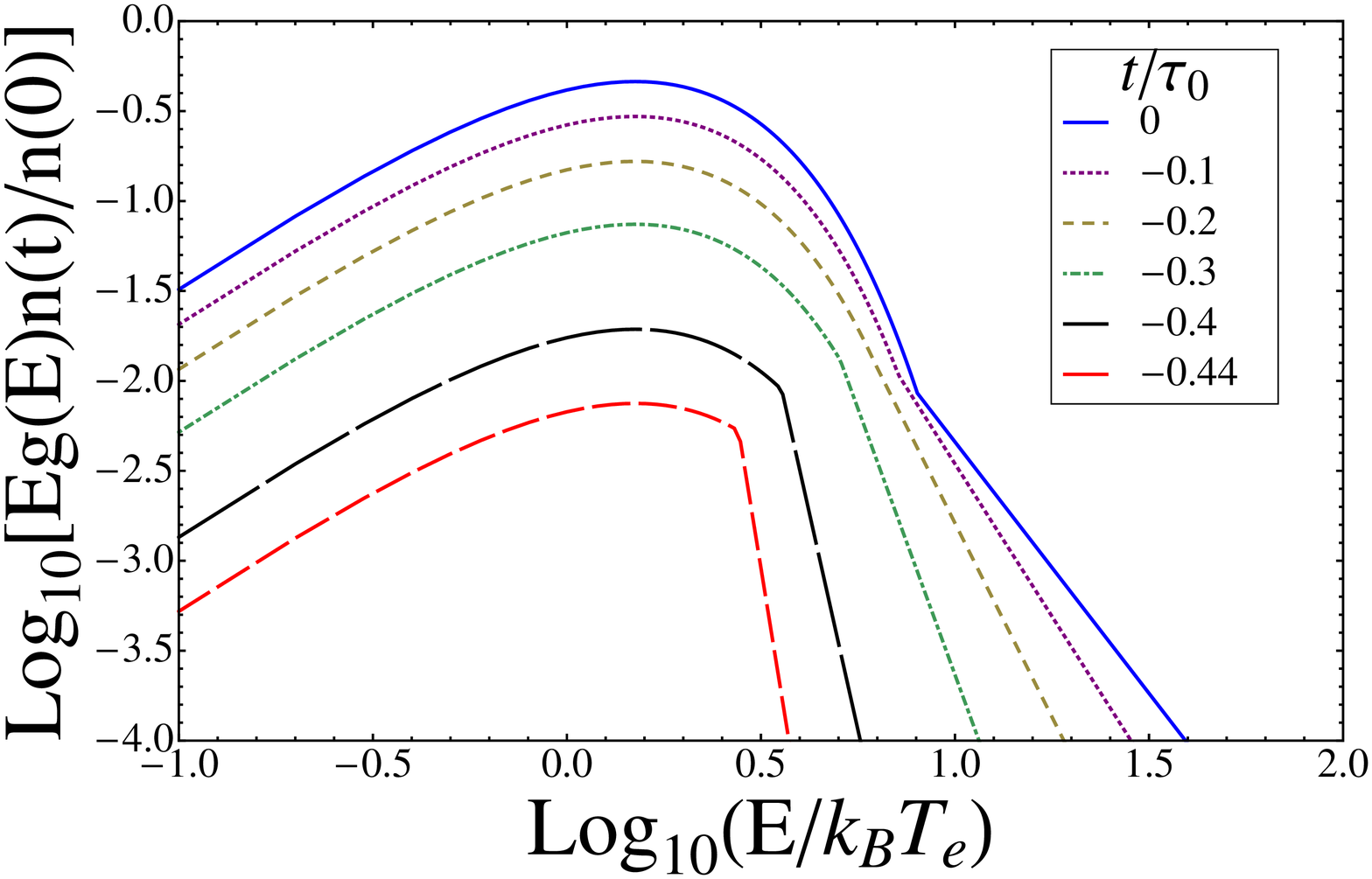}
 \includegraphics[width=5.5cm]{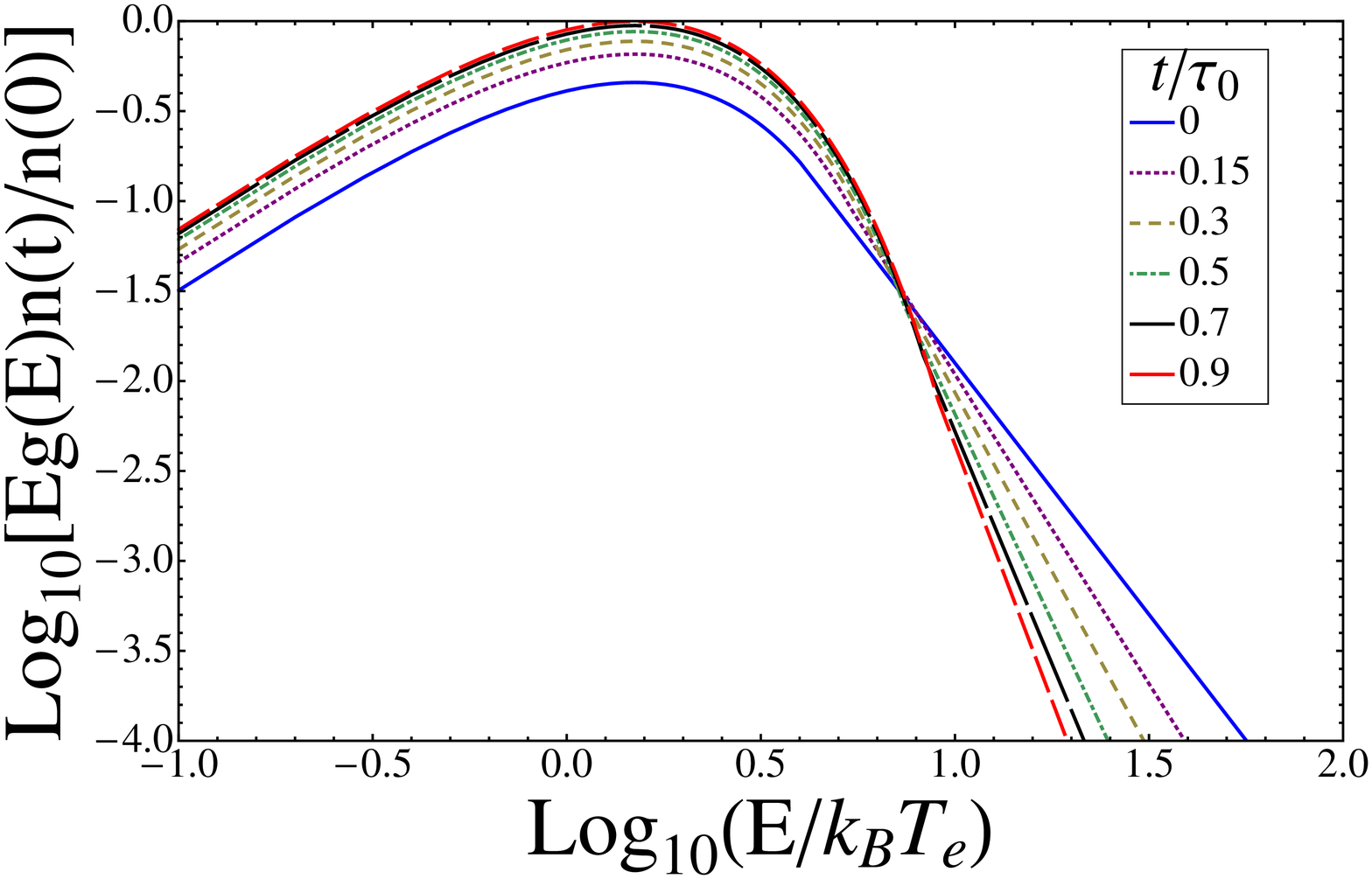}
 \includegraphics[width=5.5cm]{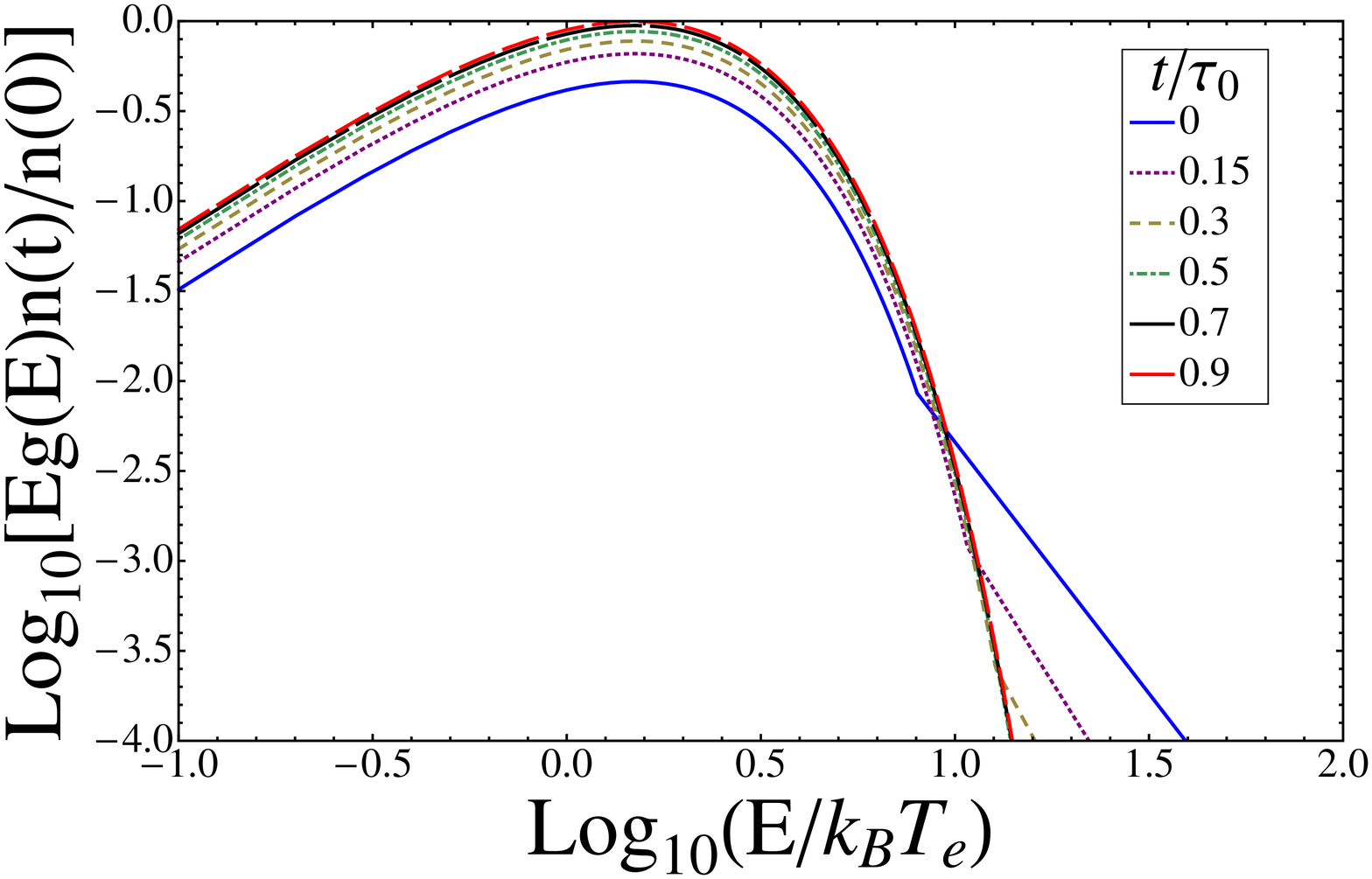}
\vspace{-0mm}
\caption{
Evolution of $g(E, t)n(t)/n_0$ during the rise
 (upper) and decay (lower) 
 phase for $r_0 = 4$ (left) and $8$ (right). The time unit is $\tau_0 =
 l_e/(b_0v_{\rm A0})$.
$C_eb_i^4 = 2b_0$, $S_a/S_s = 100$ and $\Delta=2^{1/2}b_0$ for both cases.
\label{ele}
}
\end{center}
\end{figure}

\begin{figure}[ht]
\begin{center}
 \includegraphics[width=5.5cm]{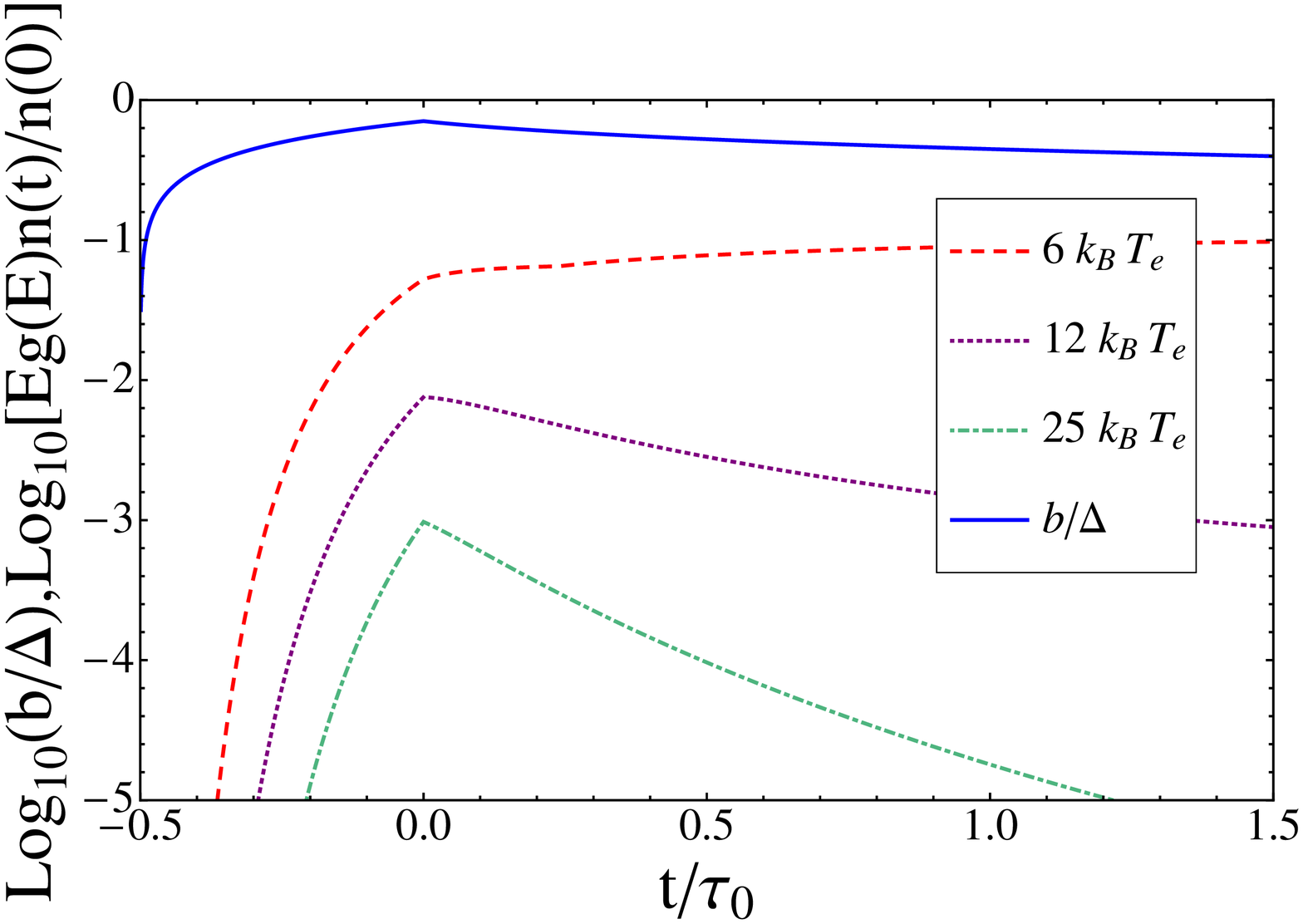}
\includegraphics[width=5.5cm]{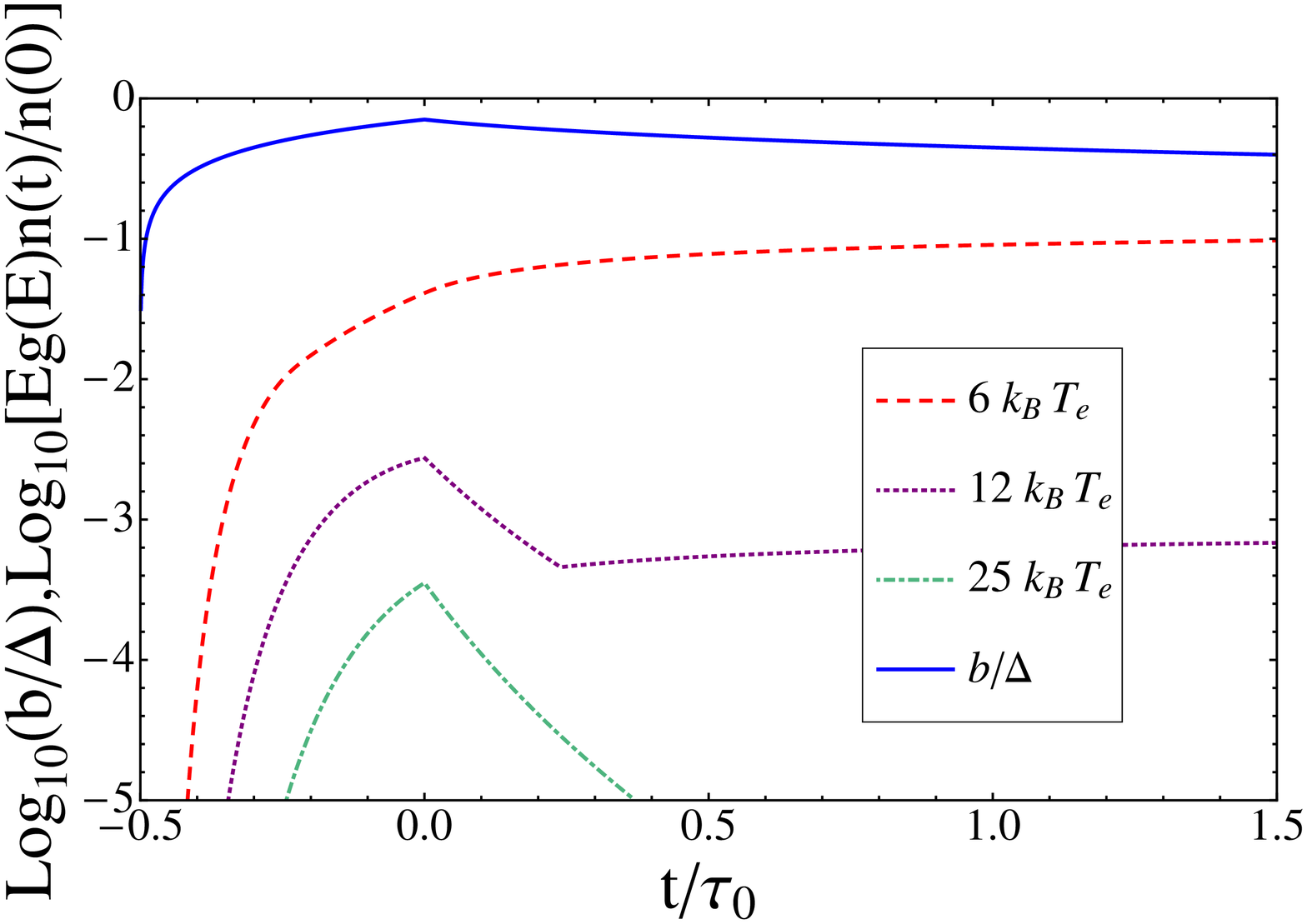}
\includegraphics[width=5.5cm]{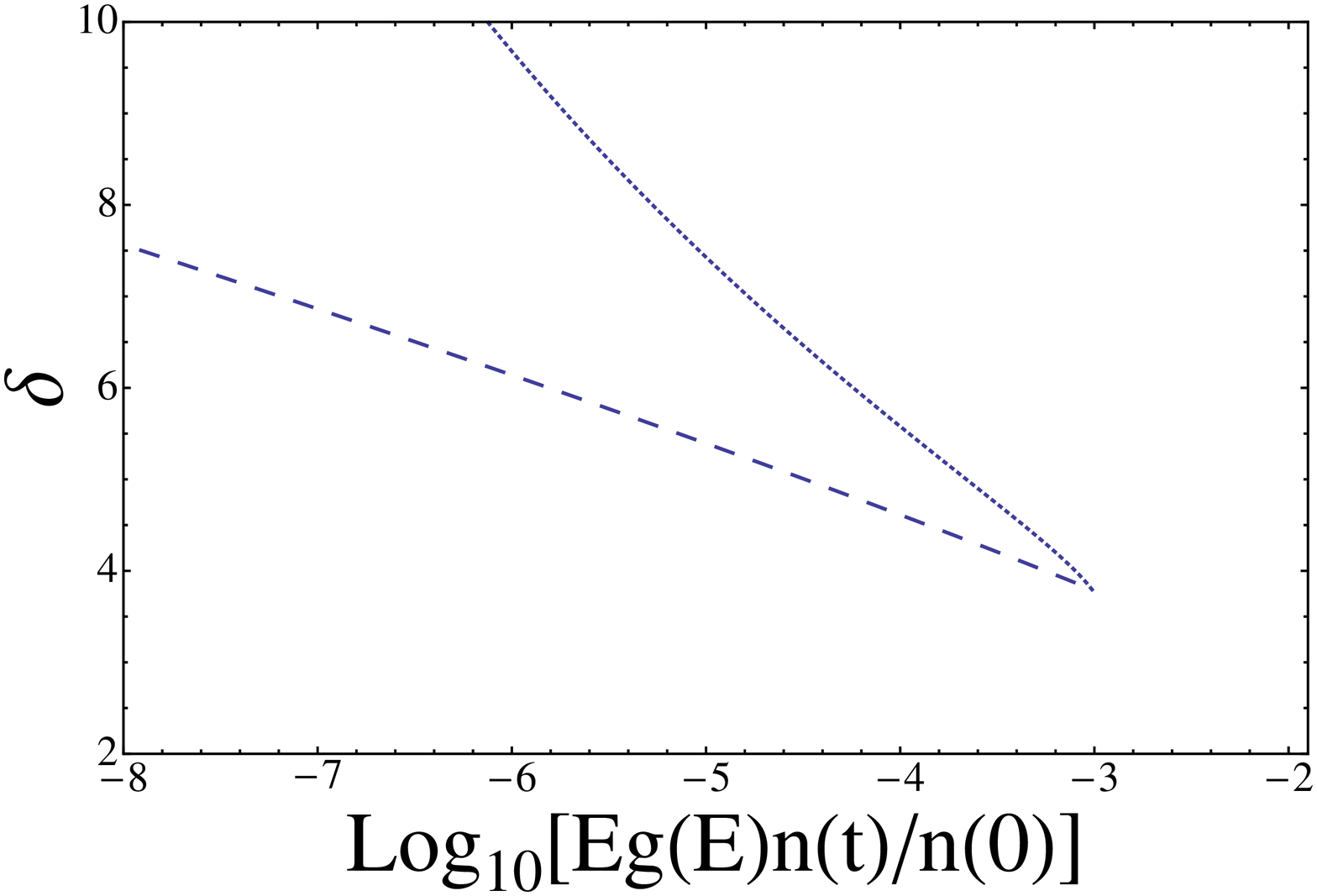}
 \includegraphics[width=5.5cm]{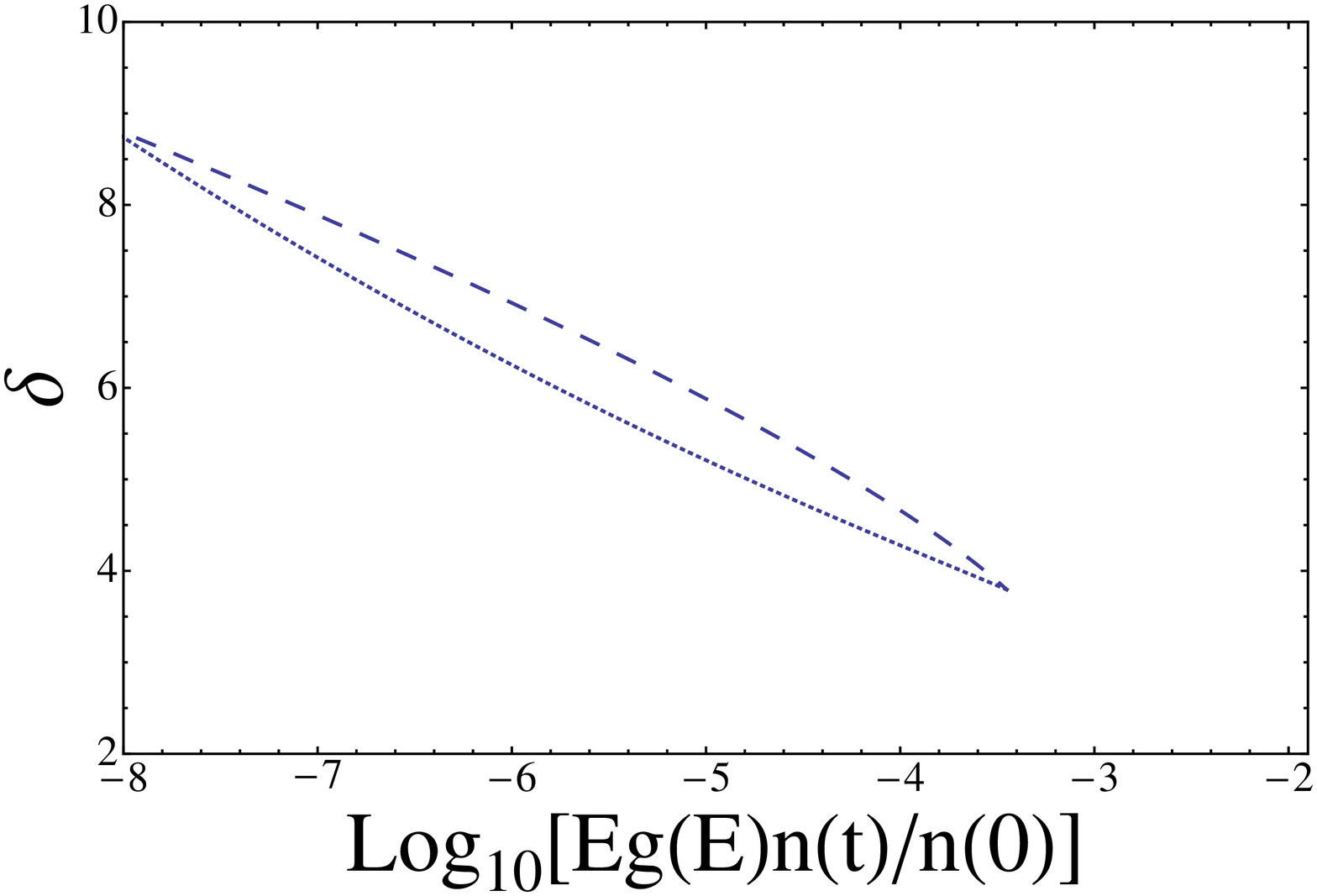}
\vspace{-0mm}
\caption{
Evolution of $g(E, t)n(t)/n_0$ (upper) and correlation between
  $\delta$ and $g(E, t) n(t)/n_0$ at $E = 25 k_{\rm B}T_e$ (lower) of the
 evaporation 
 models shown in Figure \ref{ele}. The dashed and dotted lines in the
 lower panels 
 correspond to the rise and decay phases, respectively.
\label{litele}
}
\end{center}
\end{figure} 

The solution of equations (\ref{dotd}) and (\ref{density}) critically
depends on the sign of $1-C_eb_i^4/2b_0$. 
For $C_eb_i^4=2b_0$ and given $b_0/\Delta$, $S_a/S_s$, $r_0$, $n_0$, and
$T_e(t)=T_{e0}$, $N(E, t)$ is determined for the time in units of
$\tau_0= l_e/(b_0v_{\rm A0})$. We will first study this simple
case with $T_e$ remaining constant and discuss 
more realistic applications of the model to flare observations in the
next section. For $b_0=2^{-1/2}\Delta$, and $S_a/S_s = 100$, 
Figure \ref{ele} shows the evolution of $g(E,t)n(t)/n_0$.
It is evident that, while
the electron distribution is thermal and nonthermal, respectively, at
low  ($E \ll r_0 k_{\rm B}T_e$) and
high  ($E \gg r_0 k_{\rm B}T_e$) energies, in the intermediate energy
range ($E \sim  r_0 k_{\rm B}T_e $) it experiences a transition from
nonthermal to thermal.  
Moreover, the evolution of the gradual thermal component is similar for
$r_0=4$ and $8$, in contrast to their distinct impulsive nonthermal
components.  


Figure \ref{litele} shows the time dependence of $g(E, t)n(t)/n_0$ at
several energies. In combination with Figure \ref{ele}, it is evident that,
for the ratio of transition energy to thermal energy at the turbulence peak 
$r_0 = 4$, electrons at 6 $k_{\rm B}T_e$ are
nonthermal and impulsive near the peak of $b$ and become thermal in the
gradual decay phase. At $E = 12$ and $25$ $k_{\rm B}T_e$, the distribution is
nonthermal and impulsive. 
For $r_0 = 8$, electrons at $6\ k_{\rm B} T_e$ are nonthermal in the
early rise phase and become part of a thermal distribution before the
peak of $b$. Its sharp rise near the peak of $b$ mimics the well-known
Neupert effect \citep{n68}. 
Electrons at $12\ k_{\rm B}T_e$ become thermal in the
decay phase. Compared with the model with $r_0=4$, the most distinct
feature is the much sharper decrease of $g$ at $E = 12$ and 25 $k_{\rm
B}T_e$ in the early decay phase.  
This is mostly caused by the 
relatively high values of $E_t/k_{\rm B}T_e$, which makes the density of
nonthermal electrons very sensitive to $E_t$ and therefore $b(t)$ and
$n(t)$.

\begin{figure}[ht]
\begin{center}
 \includegraphics[width=5.5cm]{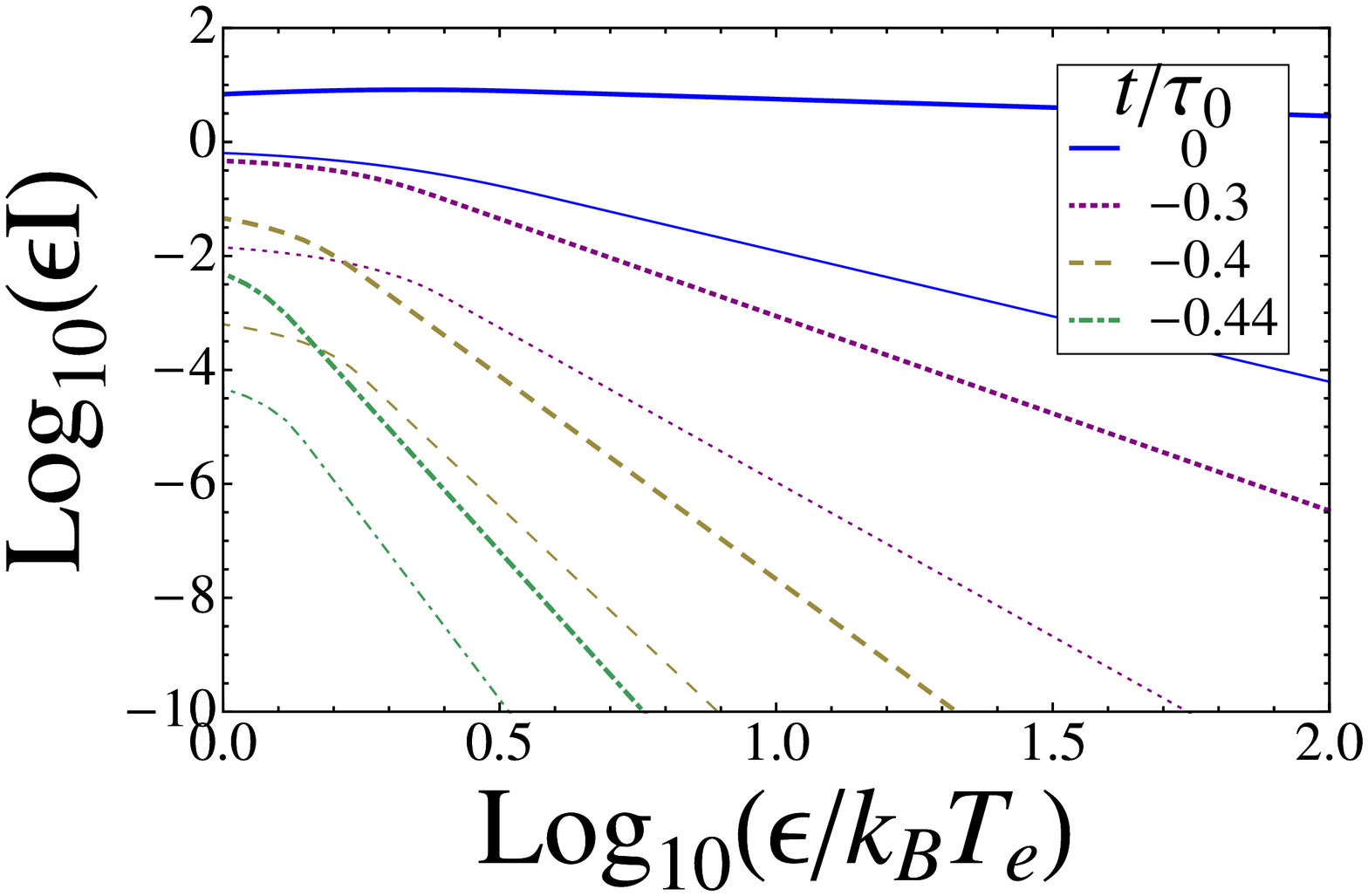}
\includegraphics[width=5.5cm]{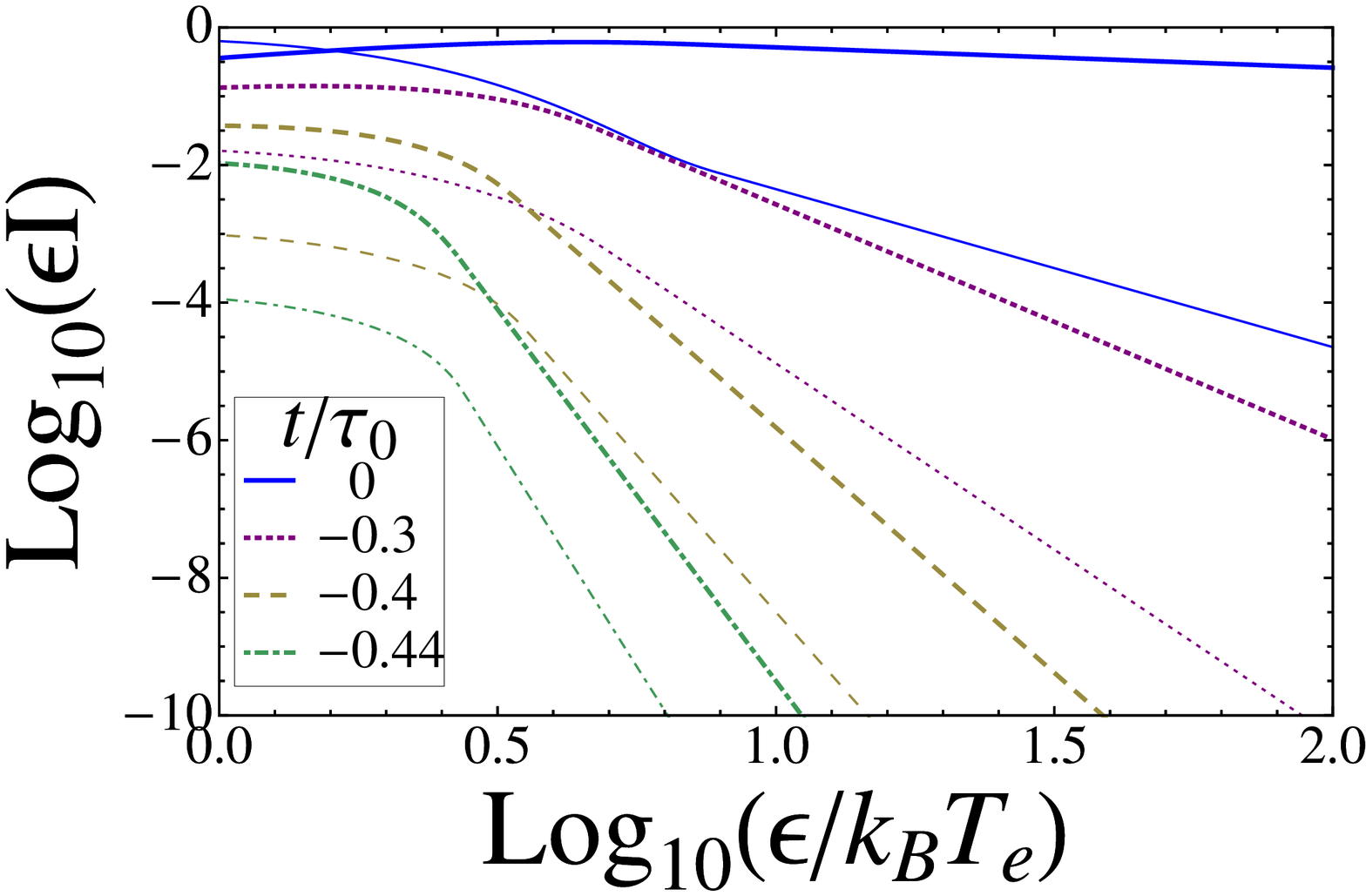}
 \includegraphics[width=5.5cm]{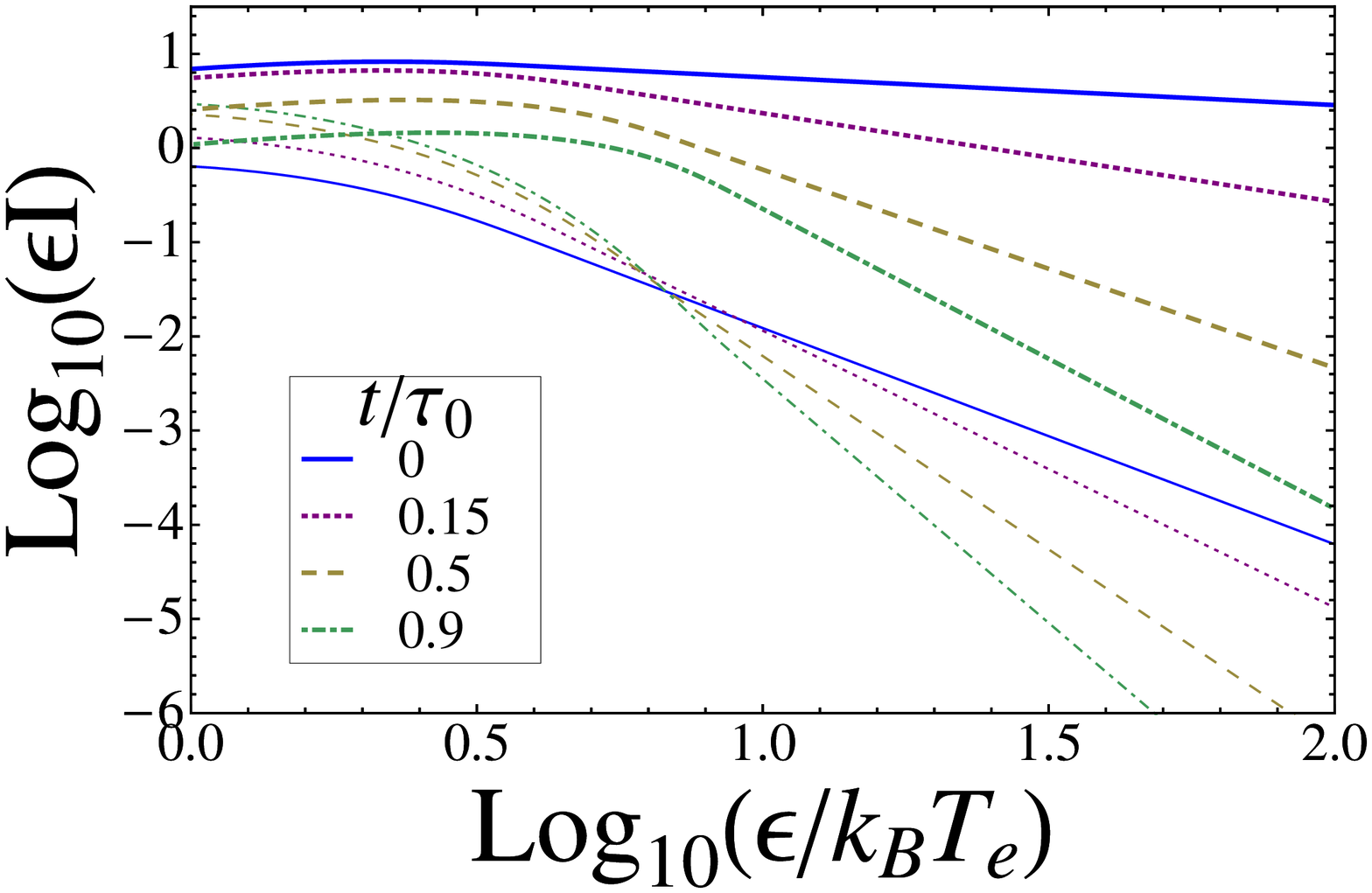}
 \includegraphics[width=5.5cm]{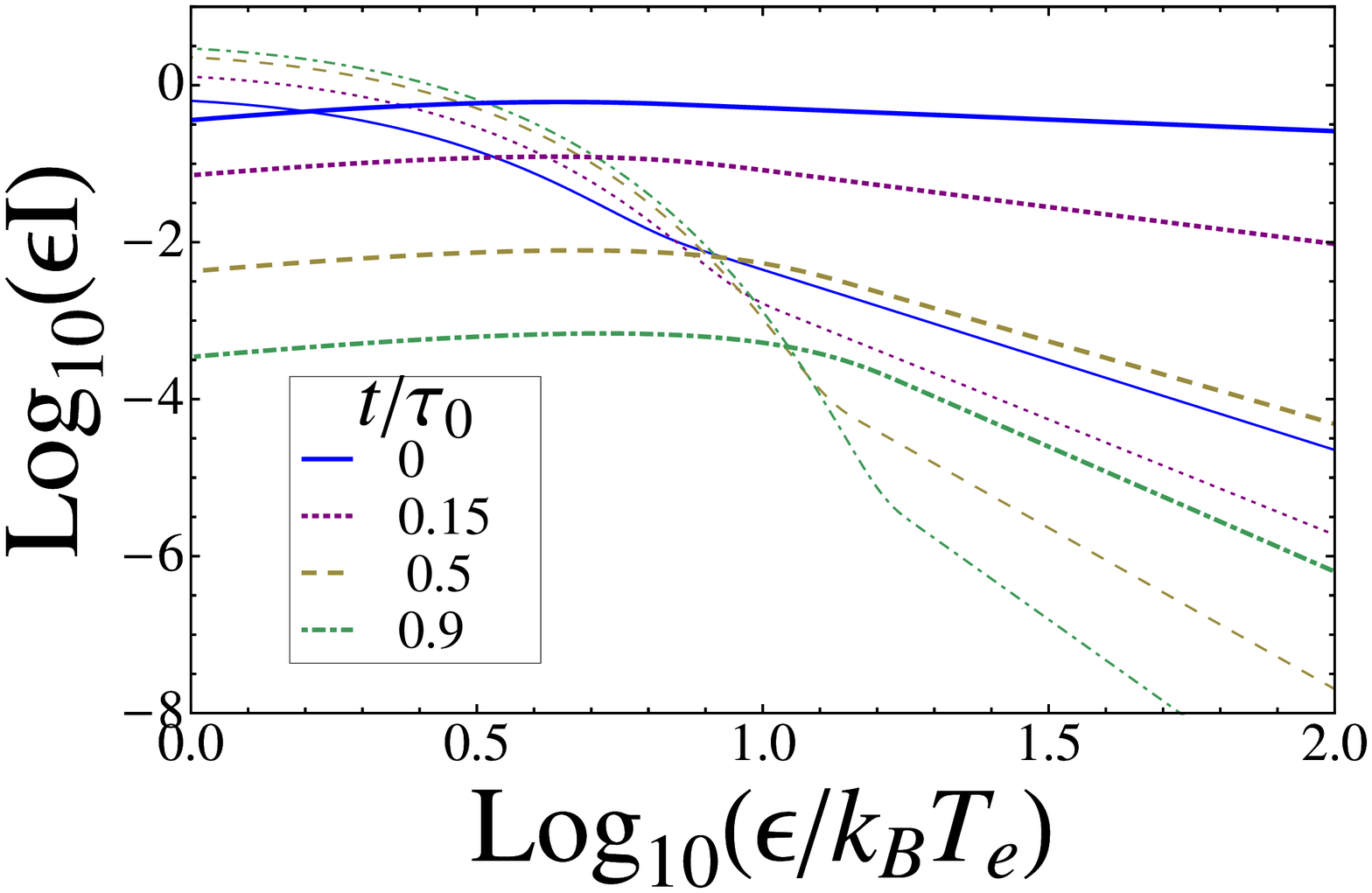}
\vspace{-0mm}
\caption{
Evolution of the energy spectrum $\varepsilon I$ during the rise
 (upper) and decay (lower) 
 phase for $r_0 = 4$ (left) and $8$ (right), where $I$
 is the specific intensity rather than the photon flux, as is more often
 used. The unit is ${2^{3/2}\alpha
 e^4l_0A n_0^2}(k_{\rm B}T_e)^{1/2}/({3{\pi}m^{3/2}c^2 R^2})$,
 where $\alpha$ is the fine structure constant and $c$ the speed of light. The
 thin and thick lines 
 are for the thin- and thick-targets, respectively.
\label{srn}
}
\end{center}
\end{figure}

Figure \ref{srn} shows the corresponding evolution of $\varepsilon
I(\varepsilon, t)$. To increase the dynamical range of the plot, we show
the energy spectrum $\varepsilon
I(\varepsilon, t)$ instead of the photon number flux spectrum  $
I(\varepsilon, t)/\varepsilon $ normally plotted in papers on specific
observations. 
The thin and thick lines are for
the first and second term of equation (\ref{I}),
and the upper and lower panels are for the rise and decay phases, respectively. 
The distinctions between the thermal and nonthermal components are less
prominent than those for electrons primarily due to the dominance of
the thick-target emission and contributions of high-energy electrons to
low-energy  X-rays. 
The nonthermal HXRs are always dominated by the thick-target component,
which dominates the rise phase of the $r_0=4$ model as well. The thin-target
component only dominates the low-energy thermal emission in the late
rise and decay phases. It is
interesting to note that a low-energy spectral flattening in the total
spectrum can result from competition between the thin- and thick-target
components, which provides an alternative to the return
current explanation for this observed phenomenon \citep{z06}. 

\begin{figure}[ht]
\begin{center}
 \includegraphics[width=5.5cm]{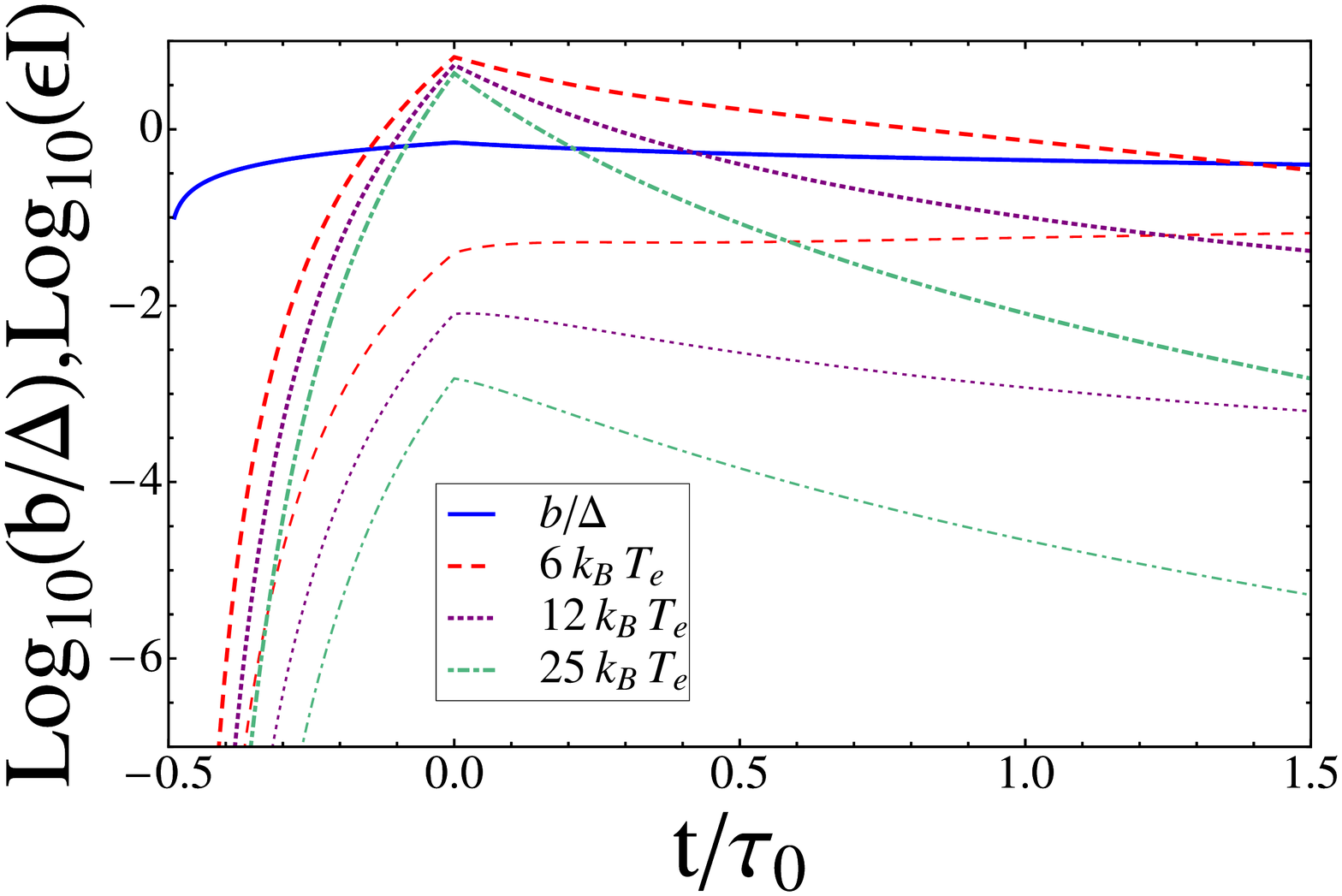}
\includegraphics[width=5.5cm]{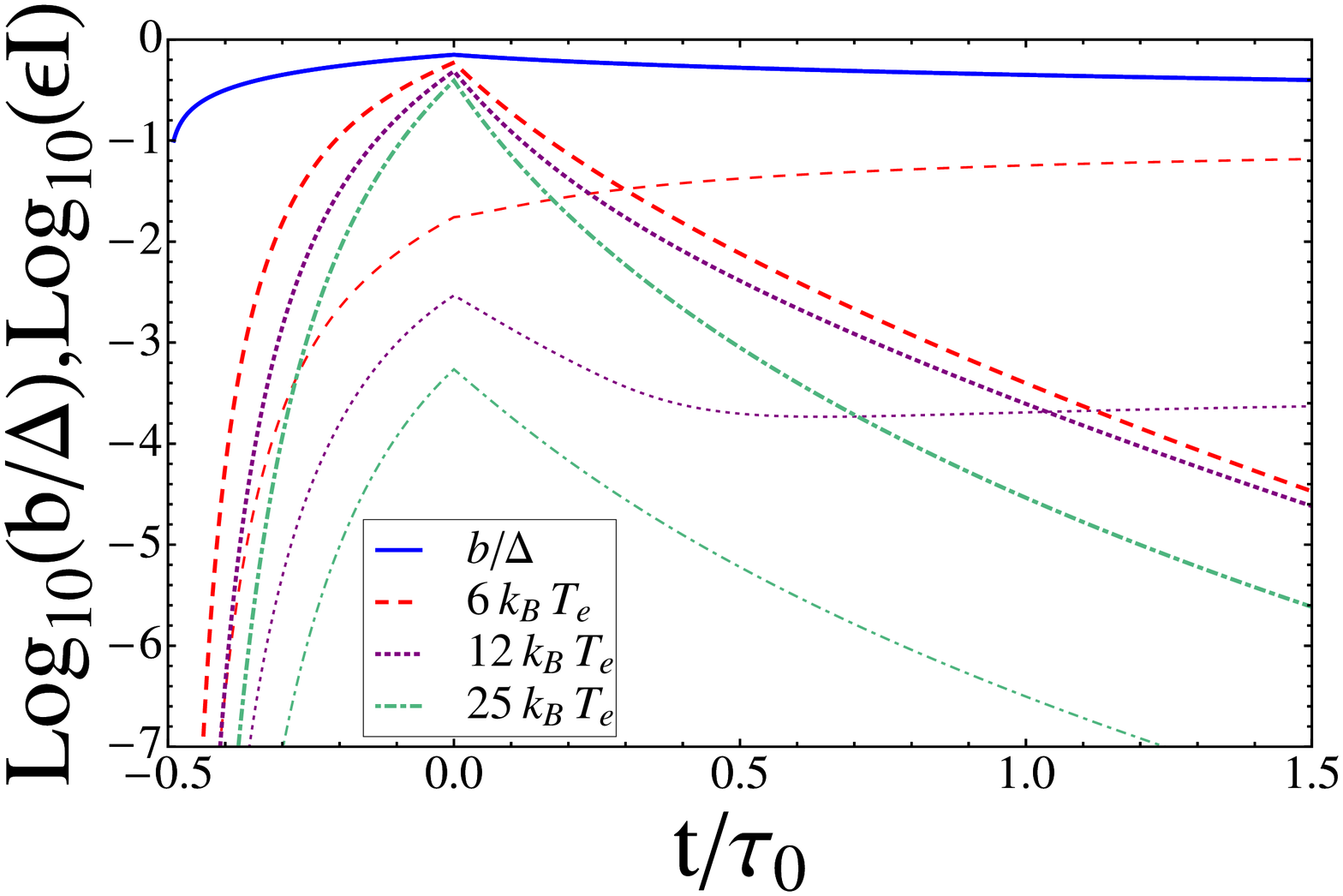}
\includegraphics[width=5.5cm]{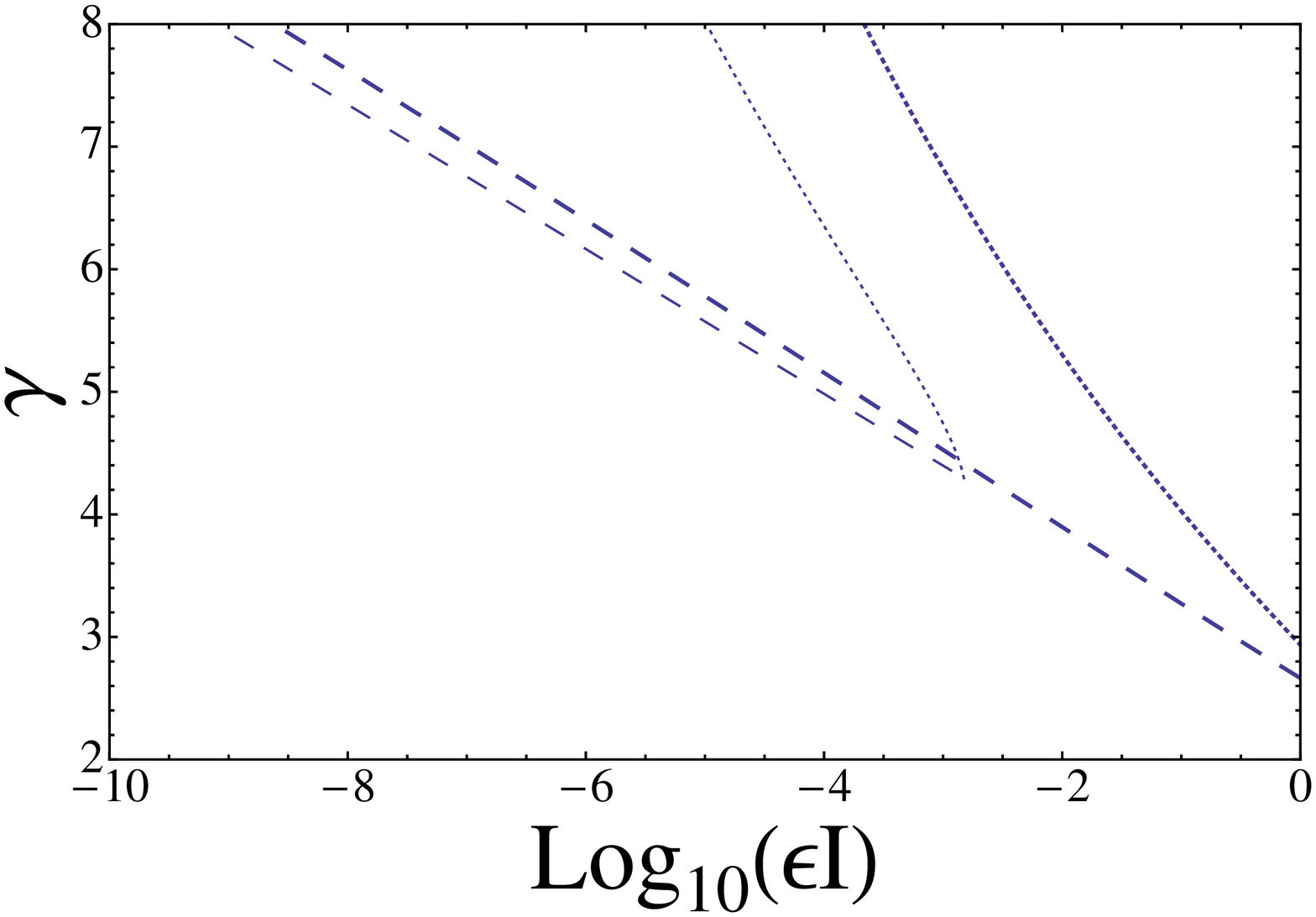}
 \includegraphics[width=5.5cm]{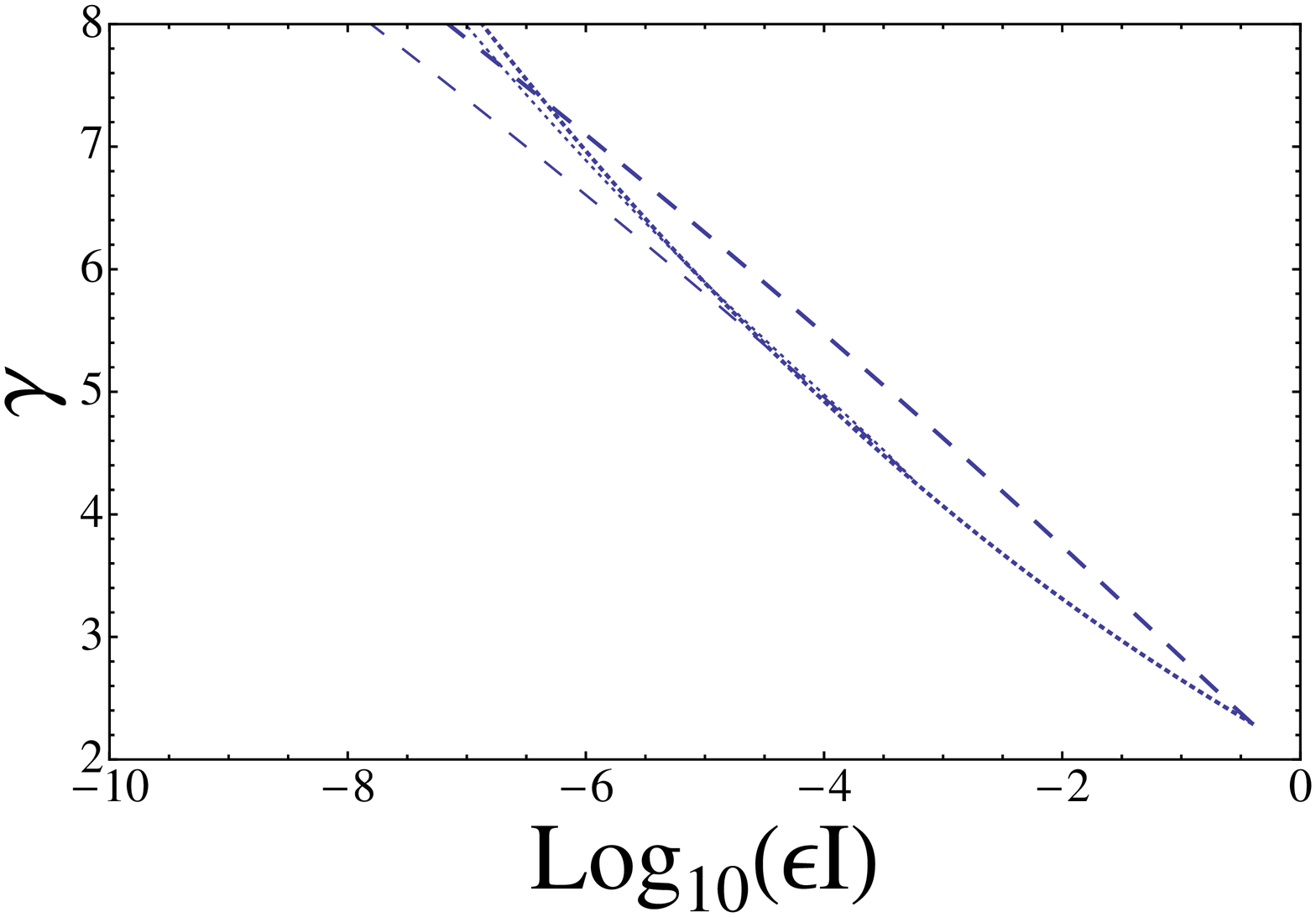}
\vspace{-0mm}
\caption{
Light curves (upper) and correlation between
 differential photon flux index $\gamma$ and 
 $\varepsilon I$ at $\varepsilon = 25 k_{\rm B}T_e$ (lower) of the evaporation
 models shown in Figure \ref{srn}. The dashed and dotted lines in the
 lower panels 
 correspond to the rise and decay phases, respectively. And the
 thin and thick lines 
 are for the thin- and thick-targets, respectively.
\label{lit1}
}
\end{center}
\end{figure} 
The upper panels of Figure \ref{lit1} show light curves of
$\varepsilon I(\varepsilon,t)$ at
several energies. 
The impulsive phase is mostly dominated by the thick-target emission and
the thin-target component dominates in the gradual phase especially at
low energies. 
Compared to the case with 
$r_0=4$, the HXR fluxes of $r_0=8$ both rise and decay more sharply
with a plateau near the peak. This plateau is less distinct at higher
energies.  
The dependence of HXRs on $r_0$ is a direct consequence of the dependence of the
nonthermal electron density on $r_0$. A high value of $r_0$
can be caused by a low temperature and/or 
high density of the background electrons for a given acceleration
timescale. The model therefore predicts 
that sharper rise and decay of HXR emission imply
lower temperature 
and/or higher 
density of the thermal component. This is the key prediction of the
model. 

Since the evolution of $\delta(t)$ is determined
by $b(t)$ (eq. [\ref{gam}]), the dependence of the HXR light curves on
$T_e$ and $n$ will affect the correlation 
between HXR flux and differential photon spectral index $\gamma =
1-{\rm d}\ln I/{\rm d}\ln\varepsilon$ frequently
observed for HXR pulses \citep{g04, b06, g06, g08}. (Note that since
$\gamma$ is usually defined in terms of photon number flux spectrum instead
of energy flux spectrum, it differs from the index of $I$ by 1.)
 The lower panels of
Figure \ref{lit1} show how the 
correlation changes from the rise to decay phase at $\varepsilon = 25
k_{\rm B}T_e$. The corresponding correlation between $\delta$ and $g(E,
t)n(t)/n_0$ 
is shown in the lower panels of Figure \ref{litele}. Since the
thick-target emission does not depend on the density of the target
region, the correlation between $\delta$ and $g(E, t)n(t)$ is similar to
the correlation between $\gamma$ and $\varepsilon I$. The thin-target
emission scales with $n(t)^2 g(E, t)$, and $\varepsilon I$ therefore decays
more slowly than $n(t) g(E, t)$.
 \citet{g04} claims
that the correlation for these pulses ``follows tendentially a slanted
V, with the rise phase forming the flatter leg''. This is in agreement
with the $r_0=4$ case. For higher values of $r_0$, our model also
predicts an opposite evolution with 
the decay phase forming the flatter leg. Such pulses do exist \citep{g08}. The
strongest pulse of the 25-Feb-2002 flare shown in Figure 4 of
\citet{g04} is one example. It is also interesting to note that this
pulse has a relatively gradual rise and sharp decay phases in agreement
with our model prediction at high energies. The majority of pulses from the
27-June-1980 flare studied by \citet{l87} also seems to be compatible with
the latter model as shown in their Figure 8. 

\section{Discussion}
\label{dis}

We have shown that the model has many characteristics similar to the observed
thermal and nonthermal X-ray emissions, especially for early impulsive flares
\citep{s07}.  It also predicts that
variations in properties of the thermal background plasma result in significant
differences in
the nonthermal particles. To test the model more quantitatively
with observations, one needs to address the
heating of electrons and ions and other important processes in flaring
loops. The evolution of $T_e$ and ion temperature $T_i$ 
depends on the
ill-understood plasma heating by turbulence and chromospheric
evaporation. 
The inhomogeneity along the loops and related production of plasma waves
will also affect the 
thermal energy gain of background particles \citep{z06, k09}. 
In cases where the energy release site is localized at the
looptop regions, one also needs to consider transport effects from
the acceleration region  to the thick-target region. Some
transport processes may modify the electron distribution by inducing
waves \citep{k09}, which may propagate into the acceleration region  and
affect both the acceleration and transport processes. The trapping of energetic
electrons in the acceleration region  also modifies the resulting photon
spectral evolution \citep{g06}.
The problem is even more complex when one
considers the effect of conduction and
radiation from coronal loops and chromospheric footpoints \citep{l09w}.
For strong evaporation, shocks may form making the large-scale bulk motion an
important energy component \citep{ly09}. 

There are already extensive numerical studies of the dynamics of
flaring loops \citep{l97, a05, k06, l09w}. Significant
uncertainties remain. The essential difficulties are rooted in the highly
dynamical nature of turbulent plasma in a strongly magnetized
environment and we still lack effective tools to address this
phenomenon in quantitative detail \citep{j09}. It is therefore not surprising 
that flare HXRs often show features over a broad scale range. For some
flares, features from the duration of the observed HXRs down to the time
resolution of the instrument can be readily identified \citep{A98}. Even
for the relatively simple early impulsive flares with the HXRs dominated
by a few distinct bursts \citep{s07, ly09}, finer structures with a
timescale much shorter than the duration of individual bursts are
expected \citep{a96}. Flare HXRs are therefore an intrinsically
complex phenomenon. Detailed deductive modeling of individual flares
is deemed to be tedious if possible, especially for extended bursts without
distinct dominant scales. 

On the other hand, for simple flares with the HXRs dominated by a few
distinct bursts, which usually follow the soft-hard-soft evolution, our
model can be readily applied to individual bursts to account for these
energetically dominant components. The energetically 
less important residuals from a successful model
fitting to specific flare observations can be attributed to small-scale
processes ignored (or averaged out) in the model. An expansion
approach in energetics therefore
may be taken to achieve more and more quantitative modeling. The
elementary energy events proposed here therefore may reflect more about the
fact that some flares are relatively simple and may be modeled in
quantitative details rather than some fundamental physical
processes. If there are universal processes operating in all flares,
these simple flares will be good candidates for us to disentangle these
processes by modeling in detail, which may lead to
interesting testable predictions. With the elementary energy release
events proposed here, 
one can carry out detailed modeling of  
early impulsive events, where properties of the background electrons 
can be readily obtained from RHESSI's high resolution observations of
X-ray emissions. Quantitative tests of the model are possible,
especially for the study of energy release during the impulsive phase
when one may assume $T_i=T_e=T_{e0}$.  

\begin{figure}[ht]
\begin{center}
 \includegraphics[width=7.5cm]{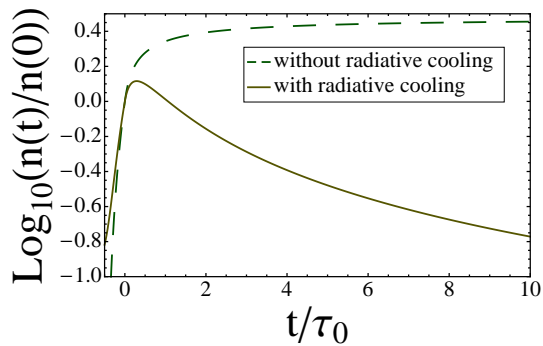}
\vspace{-0mm}
\caption{
The evolution of the average density $n$ with (solid) and without
 (dashed) the radiative cooling taken into account. See text for details. 
\label{litn}
}
\end{center}
\end{figure}

The energy loss needs to be treated properly to apply
the model to observations especially for the gradual decay phase. To
partially take into account the 
effect of energy loss during the impulsive phase, one may replace
equation (\ref{evap}) with 
\begin{equation}
\dot{n}(t) = C_en_i[\dot{\Sigma}(t)-\Lambda]/(\rho v_{\rm A}^2)\,, 
\label{evap1}
\end{equation}
where $\Lambda = \kappa_r n^2/T_e^{1/2}$ is the radiative cooling rate
and $\kappa_r \simeq 1.42\times 10^{-19}$ ergs cm$^{3}$
s$^{-1}$ K$^{1/2}$ \citep{t88}. Figure \ref{litn} shows how the density
variation is reduced due to inclusion of the radiative cooling for
$\Lambda_0/\dot{\Sigma}_0 = 0.22/b_0^4$, $C_en_ib_0^2=2n_0$, and
$b_0=1$. The density 
decrease for the result with radiative cooling considered may not
represent observations accurately because depletion of plasmas from coronal 
loops produced by flares usually occurs when the plasma temperature is
already so low that 
there are 
no significant X-ray emissions \citep{w04}. The temperature evolution must
be considered before the density reaches the peak value. This can be
done when applying the model to individual flare observations.

Flares associated with single
loops have been studied extensively by many authors \citep{v05, l06, s07, h08,
x08}. Two flares occurring on 13-Aug-2002 after 5:30 UT are also good
examples \citep{ly09}. Among these flares, there are events with the HXR
emission 
dominated by a single pulse \citep{s07}. Our elementary energy release
events can be applied to
these flares directly to study all aspects of the 
impulsive phase. These pulses usually have
a duration of less than 1 minute. It may be difficult to obtain the
detailed spectral evolution, however, one can use the model to
reproduce the observed light curves with a prescribed temperature
evolution. $T_e$, $C_e$, and $n_0$ determine the thermal X-ray light
curves. The photon spectral index at the HXR peak can be used to derive
$b_0/\Delta$, which determines the evolution of the HXR spectral index. 
The time unit $\tau_0$ can be adjusted to fit the duration of the HXR
pulse \citep{l95}.
The shape of the HXR pulse can be fitted by adjusting $S_a/S_s$ and
$r_0$.
The overall normalization $N_0$ depends on the
size and density of the looptop acceleration region. Therefore a good
fit to such 
flares would lead to well-defined model parameters. 
If observations can 
also give a reasonable measurement of the source size $l_0$, from equations 
(\ref{et}) and (\ref{Del}), one can show that the model parameters $n_0$,
$T_{e0}$, $b_0/\Delta$, $S_a/S_s$, $r_0$, $l_0$ must satisfy the following
equation 
\begin{equation}
r_0k_{\rm B}T_{e0} = (2^{5/2}\pi\ln\Lambda
 n_0l_0 S_a/S_s)^{1/2}e^2(\Delta/b_0)/[(1+2S_a/3S_s)^{1/2}-1]^{1/2}\,.
\label{cons} 
\end{equation}
$l_eS_s$ and $l_eS_a$ are introduced in equation (\ref{tac}) and can be
obtained in term of $\Delta$ with equations (\ref{Del}) and
(\ref{gam}). The magnetic field $B$ needs to be
determined from other means to get $l_e$ (also in term of $\Delta$) from
$\tau_0$ and $b_0/\Delta$. It is evident that $\Delta$ can not be
derived from the above model parameters. By replacing $\Delta$ with
$l_e$, one, however, can express $S_a$
and $S_s$ in terms of $l_e$, which are intimately related to the
energy release and 
electron interactions with the turbulence and should not change
significantly from flare to flare. Of course, if the turbulence speed
$b_0v_{\rm A}$ can also be determined through other means, one can get $l_e$
from $\tau_0$ and then derive $b_0$, $\Delta$, $S_s$, and $S_a$. 
The capability of the model to
reproduce observed light curves  with the constraint (\ref{cons}) can be
considered as a test of its 
validity. 
From fitting of a large enough sample with different peak 
HXR spectral indexes, one may obtain statistical measurements of $C_e$,
$l_e$, $b_0$, $\Delta$, $S_a$, and $S_s$. 

There are also flares with relatively long HXR pulses ($\sim 1$
min), e.g., the 20-Sep-2002 flare starting at 9:23 UT and the flare
studied in detail by 
\citet{l06}. Given the relatively gradual HXR peak of these events, the
elementary energy release events will not produce a good fit
to the light curves. For these events, one can obtain detailed spectral
evolution. Equation (\ref{gam}) therefore can be used with the observed
$\delta(t)$ to derive the evolution of $b(t)$ and $\Sigma(t)$ during the
HXR pulses. The 
temperature evolution $T_e(t)$ can also be obtained from the
spectral evolution. With $T_e(t)$ and $\delta(t)$ as inputs, one can
again fit the observed light curves with the model to test its
validity, especially the evaporation model described with equation
(\ref{evap}).  
In a more accurate treatment, one should replace equation
(\ref{evap}) with an appropriate energy conservation equation to have
more insights of the evaporation and energy release processes.
For more general cases, where there can be a preheating phase
\citep{b09} and/or many HXR pulses \citep{v05}, suggesting continuous
energy release, one can treat $b(t)$ as an input and apply the model to
observations to derive its evolution, which can be used for further
study of plasma heating, particle acceleration, and chromospheric
evaporation.

\section{Conclusions}
\label{con}

Although the particle acceleration timescale is short, the energy
release processes of solar flares are multi-scale phenomena and some
long timescale processes may have imprints on the characteristics of
nonthermal components. In this paper, we demonstrate how the evolution of
X-ray emissions may be affected by the chromospheric evaporation
in the context of stochastic acceleration by turbulent electromagnetic
fields. We 
consider the simplest scenario, where the energy release in closed loops
is described with a turbulence intensity $b_0 B$ and generation scale
$l_e$. It is shown that the HXR
flux has a sharper decay phase for 
events with lower temperature and/or higher density of the background
electrons, which makes the transition energy between the thermal and
nonthermal component much higher than $k_{\rm B}T_e$. 
As the ratio, $r_0$, of the transition energy to the thermal energy at
the HXR peak is
varied from low to high values (by varying the density and
temperature) the HXR behaviour changes between two types of spectral
evolution. If this ratio is low, the flux change in the rise phase is
higher than in the decay phase for a given change in the spectral index
(the commonly observed behaviour) and vice versa if $r_0$ is high --- a
behaviour which is also observed, though less frequently.

Solar flares involve many physical processes that give rise to significant
uncertainties in our understanding of the plasma heating and particle
acceleration. Even for relatively simple flares associated with single
loops, the mass, momentum, and energy transport along the loops during
the impulsive phase can be highly dynamical and complicated. It is therefore
challenging to study the flare energy release with all the relevant
processes treated properly. The problem can be simplified significantly
if we are mostly concerned with the energetically dominant processes
such as the energy release, particle
acceleration, and chromospheric evaporation averaged over a relatively
long timescale of observational interest. We discuss how the model may
be applied to RHESSI 
observations of flares  for
quantitative investigations, especially for early impulsive flares with the
HXRs dominated by a single pulse with soft-hard-soft spectral evolution. 

While the short acceleration timescale may suggest microscopic processes,
the large amount of energy released during flares reveals a predominantly
macroscopic process. One critical aspect of solar
flare studies is to understand the connection between the microscopic
particle acceleration and the macroscopic energy release. The
latter has been studied with magnetohydrodynamical simulations. 
The
study of particle acceleration is limited to a few very specific
mechanisms. Although
the high
degree of freedom on large scales (in connection with the magnetic field
configuration and initial and boundary conditions) inevitably lead to
multi-scale features giving rise to significant complexities, the small
scale processes are expected to follow some patterns as least in
a statistical sense. One therefore may focus on flares with relatively
simple HXR light curves. The model presented here makes it possible to
infer the 
energy release rate from the observed HXR emission for these flares. It
is therefore 
possible to probe the macroscopic energy release processes with the
impulsive nonthermal emission. 
Successful applications of the model to
flare observations will lead to quantitative measurements of the particle
acceleration and scattering rate by turbulent electromagnetic fields.
The obtained $b(t)$ can be compared
with MHD simulations of flares or models of large scale energy
release processes for more quantitative and self-consistent modeling. 

\acknowledgements
This work is supported by the EU's SOLAIRE
Research and Training Network at the University of Glasgow
(MTRN-CT-2006-035484), the Ministry of Science and Technology of China
(Grant No.2006CB806302) and by Rolling Grant ST/F002637/1 from the UK's
Science and Technology Facilities Council. 

{}

\end{document}